\documentclass[aps,pre,onecolumn]{revtex4-1}
\usepackage{amsmath, amssymb}
\usepackage[small]{caption}
\usepackage{graphicx,dblfloatfix}
\usepackage{geometry}
\usepackage{hyperref}
\usepackage{mathtools}
\usepackage{physics}
\usepackage{relsize}
\usepackage{subcaption}
\usepackage[inline]{enumitem}

\graphicspath{{./figs/}{.}}

\newcommand*\eq[1]{{#1}^{\left(\mathrm{eq}\right)}}

\begin{document}
	\title{A Partial Entropic Stabilization of Lattice Boltzmann MHD}
	\author{Christopher Flint}
	\author{George Vahala}
	\affiliation{Department of Physics, College of William \& Mary, Williamsburg, Virginia 23185}
	\date{\today}
	
	\begin{abstract}
		The entropic lattice Boltzmann algorithm of Karlin et. al. is partially extended to magnetohydrodynamics, based on the Dellar model of introducing a vector distribution for the magnetic field.  This entropic ansatz is now applied only to the scalar particle distribution function so as to permit the many problems entailing magnetic field reversal.  A 9-bit lattice is employed for both particle and magnetic distributions for our two dimensional simulations.  The entropic ansatz is benchmarked against our earlier multiple relaxation lattice-Boltzmann model for the Kelvin-Helmholtz instability in a magnetized jet.  Other two dimensional simulations are performed and compared to results determined by more standard direct algorithms:  in particular the switch over between the Kelvin-Helmholtz/tearing mode instability of Chen et. al., and the generalized Orszag-Tang vortex model of Biskamp-Welter.  Very good results are achieved.
		
	\end{abstract}
	\maketitle
	\section{Introduction}\label{Sec:Intro}
	
	The lattice Boltzmann (LB) algorithm has proven to be an extremely interesting method for the solution of Navier-Stokes \cite{succiBook} flows because of its simplicity, extreme parallelizability and accuracy.  Even though it is technically second order accurate it appears more comparable in accuracy to the pseudo-spectral computational methods.  One of the major constraints on LB is that it is prone to numerical instability in certain parameter regimes:  there is no inherent mechanism to enforce the LB distribution function to remain non-negative in time, particularly in strong turbulence simulations.
	
	This has led to a generalization of the simple single-relaxation-time (SRT) LB collision operator to its multiple-relaxation-time (MRT) cousin \cite{MRT11,MRT12,MRT13}.   With these extra degrees of freedom one can achieve greater numerical stability, but the choice of these extra parameters is problem dependent, not known a prior, and can influence the fluid viscosity coefficient (and thus the associated Reynolds numbers).  An alternate approach to achieving numerical stability is through an entropic principle and a discrete H-theorem
\cite{entropic1,entropic2,karlinIntro2,boghosianIntro1,boghosianIntro2,boghosianIntro3,keatingIntro1,keatingIntro2,flint9bit,flintMRT,DellarBulkViscosity,DellarMRT,DellarMoments,premnath2009,Bosch,pattison,riley}.  In some respects, the entropic approach can be viewed as an optimal subset of MRT algorithms in which emphasis is placed on an algebraically determined entropy stabilizing parameter that is not directly dependent on the MRT collisional rates and which does not affect the fluid viscosity.

   Here we partially extended these entropic ideas to magnetohydrodynamics (MHD).  
   LB was first extended to MHD by Succi \emph{et.\ al.} \cite{MHD1st} following on the heels of a cellular automata approach of Montgomery and Doolen \cite{Montgomery}.
   An important breakthrough to a first principle LB-MHD model was by Dellar \cite{DellarMHDLB} who introduced a vector kinetic equation for the magnetic distribution function.  In conventional LB for Navier-Stokes, the zeroth moment of the scalar distribution function yields the density $\rho$, while the fluid velocity $\vec u$, is retrieved from the first moment.  In the Dellar model for LB-MHD, the zeroth moment of the vector magnetic distribution function yields the magnetic field $\vec B$ itself.  Not only does this permit moment closure at a lower level than in Navier-Stokes but it yields a consistent discrete approximation  to $\div \vec B = 0$ to machine round-off error \cite{DellarMHDLB}.  However there is a significant difference between Navier-Stokes and MHD:  in MHD there are many very important applications where there is a reversal in the magnetic field $\vec B$.  Hence a first principle extension of an entropic principle, which relies on maximization of a concave function, cannot be applied to the magnetic field vector distribution function.  Nevertheless, we shall apply the entropic stabilization scheme of Karlin et. al. \cite{entropic1} to the scalar distribution function, and no such constraint on the vector distribution function.  While some may insist that we are thus not forming any entropic stabilization as such, we will find that our partial entropic stabilization (as we shall call it) does permit numerically stable simulations at arbitrary small viscosity.  This can not be achieved even in MRT LB-modeling, which only has static relaxation rates.   We also shall show, from several different simulations, that there seems to be an increased stabilization in the LB-MHD algorithm due to this partial entropy constraint on the scalar particle distribution function.  This can be attributed to the effect of the partial entropic parameter on the magnetic field because of the magnetic field coupling that exists within the velocity momentum equation/Navier-Stokes equation  

	In Section \ref{Sec:Moments}, we present a moment-based representation for LB-MHD, while our partial entropic algorithm is outlined in Sec. \ref{Sec:Entropic}.  In Sec. \ref{Sec:Simulation} we present some two dimensional (2D) simulation results of our partial entropic LB-MHD algorithm:  magnetic reconnection in the Kelvin-Helmholtz and the magnetic tearing instability as well as on the Biskamp-Welter profile for 2D MHD. We have concentrated on 2D MHD because of its much lower computational costs as compared to 3D LB-MHD.  This is appropriate since 2D and 3D MHD there is a direct cascade of energy to small spatial scales - unlike 2D Navier-Stokes turbulence which exhibits an inverse cascade of energy to larger and larger spatial scales.  Our partial entropic stabilization algorithm is readily extended to 3D.

	\section{Moment Basis Representation for  Multiple Relaxation Model for LB-MHD}\label{Sec:Moments}
\begin{figure}[h]
		\centering
		\includegraphics[width=2.25in, keepaspectratio=true]{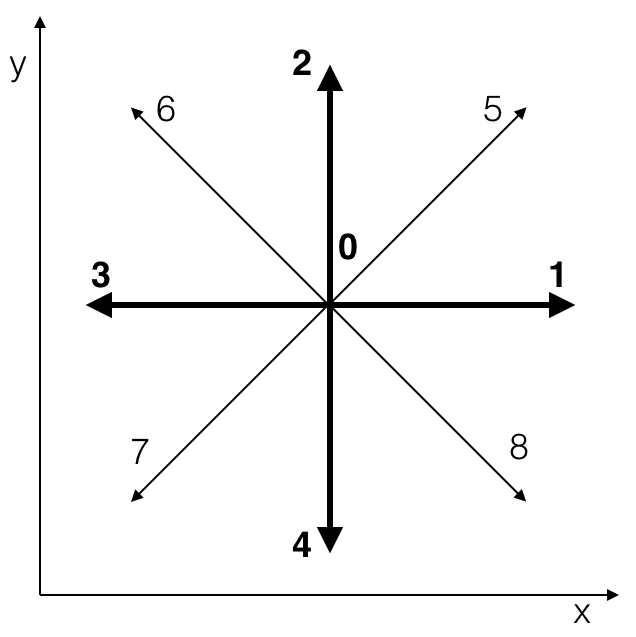}
		\caption[.]{\label{LatticeRep} The kinetic lattice vectors for 2D LB-MHD are, in our 
			$D2Q9$ model,  $\vec c_i = \left(0,0\right),\left(0,\pm 1\right),\left(\pm 1,0 \right) \ ,\left(\pm 1,\pm 1\right) \ $.  $w_i$ are appropriate weight factors dependent on the choice of lattice:  $w_0 = \frac{4}{9}$; for speed 1, $w_i = \frac{1}{9}$; and for speed  $\sqrt 2$ , $w_i = \frac{1}{36}$
			.}
		
	\end{figure}

   There are quite a few MRT extensions \cite{flint9bit,DellarMoments,pattison,Dellar2011} of the original SRT LB-MHD model of Dellar \cite{DellarMHDLB}.  However, for simplicity, we shall work with only an SRT model for the vector magnetic field distribution $\vec g_k$, and an MRT model for the scalar distribution function $f_i$, where the subscripts denote the velocity streaming directions


	\begin{eqnarray}
	& \label{LBKinEqn}\left( \partial_t + \partial_\gamma c_{\gamma i} \right) f_i = \sum_j X^{'}_{ij} \left( \eq f_j - f_j \right)  \\
	& \label{LBMagEqn}\left( \partial_t + \partial_\gamma c_{\gamma k} \right) \vec g_k = Y^{'} \left( \vec g_k^{\,(\mathrm{eq})} - \vec g_k \right)
	\end{eqnarray}
	with the moments 
	\begin{gather}
	\begin{array}{ccccc}
	\sum_i f_i = \rho &,&  \sum_i f_i  \vec c_i = \rho \vec u \quad &\mathrm{and} & \sum_k \vec g_k = \vec B
	\end{array}
	\end{gather}
	It is convenient to employ the summation convention only over the Greek indices which give the vector nature of the fields ($\gamma =1,2$ for 2D), while the Roman indices run over the
	corresponding (kinetic) lattice vectors  $\vec c_i, i = 0.. 8$ for the 9-bit model in 2D (see Fig. \ref{LatticeRep}).   Summation over the Roman indices will always be made explicit.  $X^{'}_{ij}$ is the MRT collision operator for the evolution of $f_i$ while $Y^{'}$ is the SRT for the evolution of $\vec g_k$.  The MHD viscosity and resistivity transport coefficients are determined from these kinetic relaxation rates.
	
	It is well known that the minimal LB representation of MHD equations on a square lattice is a 9-bit velocity streaming for $f_i$ and just 5-bit streaming for $\vec g_k$.  This is because $\vec u$ is defined from the first moment of $f_i$ while $\vec B$ is defined as the zeroth moment of $\vec g_k$.  It is convenient (and helpful for numerical stability) to employ the 9-bit streaming model for both kinetic equations.  To recover the MHD equations in the Chapman-Enskog limit of the (discrete) kinetic equations, we take the well-known choice of relaxation distribution functions  $f{_i}^{(eq)}$ and $\vec g_k^{\,(eq)}$ 

	\begin{eqnarray}
	&\label{feq} \eq f_i = w_i \rho \left[ 1 + 3\left( \vec c_i \cdot \vec u \right) + \frac{9}{2}\left( \vec c_i \cdot \vec u \right)^2 - \frac{3}{2} \vec u^{\,2} \right] + \frac{9}{2} w_i \left[ \frac{1}{2} \vec B^2 \vec c_i^{\,2} - \left( \vec B \cdot \vec c_i \right)^2 \right]  , i = 0, .. ,8 \\
	& \vec g_k^{\,(\mathrm{eq})} = w_k \!\left[ \vec B + 3 \left\lbrace \left( \vec c_k \cdot \vec u \right) \vec B - \left( \vec c_k \cdot \vec B \right) \vec u \right\rbrace \right] , k = 0, .., 8
	\end{eqnarray}
	
	
	The LB-MHD equations are typically solved by an operator-splitting method with time advancement coming from streaming and collisional relaxation.  
	The excellent parallelization of the LB-MHD  algorithm is now apparent: the discrete kinetic equations are solved using the streaming operation which is a simple shift of the data from one lattice point to another, while the collision step is a purely local operation with its evaluation requires only data from only that grid point.  The streaming requires MPI only when the shift has data moving from one processor domain to another -- and this can also be well parallelized.
	What makes LB so attractive is that the computationally difficult nonlinear convective derivatives $\vec u \cdot \nabla \vec u$, $\vec u \cdot \nabla \vec B$, $\vec B \cdot \nabla \vec u$ and $\vec B\cdot \nabla \vec B$ are replaced at the lattice kinetic level by simple linear advection and polynomial nonlinearities in the relaxation distributions.
	
	In MRT-LB it is natural to perform the collisional relaxation in moment space (because of the local conservation of mass and momentum constraints) and the streaming in the distribution space $f_i, \vec g_k$.  There is a 1-1 map between these spaces.  For the moment basis it is obvious to include the conservation moments  (the zeroth and first moments of the $f_i$ and the zeroth moment of $\vec g_k$), while the remaining higher moments are somewhat arbitrary \cite{lallemand1,lallemand2}.   In particular, we consider the same constant $9 \cross 9$  $\mathrm T$- matrix that connects the scalar distributions $(f_i, i=0..8)$ to their moments $(M_i, i=0 .. 8)$ as for the vector magnetic distributions $(\vec g_k, k = 0 .. 8)$ with their moments $(\vec N_k, k = 0 .. 8)$
	\begin{equation}\label{Transformation}
	\begin{array}{ccc}
	M_i = \sum_{j=0}^8 \mathrm T_{ij}f_j &, & \vec N_{k} = \sum_{q=0}^8  \mathrm T_{kq} \vec g_{q}
	\end{array}
	\end{equation}
	with 
	\begin{equation}\label{FluidTMatrix}
	\mathrm {T}=\left( \def\arraystretch{1.2} \begin{array}{c}
	\boldsymbol{1} \\ 
	c_x \\ 
	c_y \\ 
	c_xc_y \\ 
	c^2_x \\ 
	c^2_y \\ 
	c_x^2 c_y \\ 
	c_x c_y^2\\ 
	c_x^2 c_y^2 \end{array}
	\right)=\left( \def\arraystretch{1.2} \begin{array}{ccccccccc}
	1 & 1 & 1 & 1 & 1 & 1 & 1 & 1 & 1 \\ 
	0 & 1 & 0 & -1 & 0 & 1 & -1 & -1 & 1 \\ 
	0 & 0 & 1 & 0 & -1 & 1 & 1 & -1 & -1 \\ 
	0 & 0 & 0 & 0 & 0 & 1 & -1 & 1 & -1 \\ 
	0 & 1 & 0 & 1 & 0 & 1 & 1 & 1 & 1 \\ 
	0 & 0 & 1 & 0 & 1 & 1 & 1 & 1 & 1 \\ 
	0 & 0 & 0 & 0 & 0 & 1 & 1 & -1 & -1 \\ 
	0 & 0 & 0 & 0 & 0 & 1 & -1 & -1 & 1 \\ 
	0 & 0 & 0 & 0 & 0 & 1 & 1 & 1 & 1 \end{array}
	\right)
	\end{equation}
	where the Cartesian components of the corresponding $9$-dimensional lattice vectors are just
	\begin{equation}
	\begin{array}{cccc}
	c_x=\left\{0,1,0,-1,0,1,-1,-1,1\right\} &, &c_y=\{0,0,1,0,-1,1,1,-1,-1\} &.
	\end{array}
	\end{equation}                              
	For the scalar distributions, the 1\textsuperscript{st} row of the $\mathrm T$-matrix is just the conservation of density while the 2\textsuperscript{nd} and 3\textsuperscript{rd} rows are just the conservation of momentum (2D).  For the vector magnetic distributions the 1\textsuperscript{st} row of the $\mathrm T$-matrix is the only collisional invariant.
	
	With this moment basis, the MRT collisional relaxation rate tensor $X^{'}_{ij}$ is diagonalized with the $T-$ matrix as a similarity transformation.  It is convenient to denote this diagonal matrix with elements $X_{i} \delta_{ij}$.
In the $D2Q9$ phase space, the relaxation rate $X_j$ is associated with the corresponding moment $M_j$, $j=0..8$.  Similarly for the magnetic distributions in SRT, there is just a single collisional relaxation rate for each magnetic moment $\vec N_k$, and this will be denoted by  $Y$.  
	
	In particular, the equilibrium moments can be written in terms of the conserved moments:
	\begin{align}\label{KinMomentEq}
	\begin{split}
	\arraycolsep=15pt
	\begin{array}{lll}
	\eq M_0 = M_0 = \rho & \eq M_1 = M_1 = \rho u_x & \eq M_2 = M_2 = \rho u_y 
	\end{array}
	\\
	\arraycolsep=10pt
	\begin{array}{ll}
	\eq M_3 = \rho u_x u_y - B_x B_y &
	\eq M_4 = \frac{1}{6} \left( 6 \rho u_x^2 + 2 \rho - 3 \left( B^2_x - B^2_y \right) \right) \\
	\eq M_5 = \frac{1}{6} \left( 6 \rho u_y^2 + 2 \rho + 3 \left( B^2_x - B^2_y \right) \right) &
	\eq M_6 = \frac{1}{3} \rho u_y \\
	\eq M_7 = \frac{1}{3} \rho u_x &
	\eq M_8 = \frac{1}{9} \rho \left( 1 + 3 u_x^2 + 3 u_y^2 \right) \\
	\end{array}
	\end{split}
	\end{align}
	\begin{equation}\label{MagMomentEq}
	\arraycolsep=5pt
	\begin{array}{lll}
	\eq N_{\alpha 0} = N_{\alpha 0} = B_\alpha & 
	\eq N_{\alpha 1} = u_x B_\alpha - u_\alpha B_x &
	\eq N_{\alpha 2} = u_y B_\alpha - u_\alpha B_y \\
	\eq N_{\alpha 3} = 0 &
	\eq N_{\alpha 4} = \frac{B_\alpha}{3} &
	\eq N_{\alpha 5} = \frac{B_\alpha}{3} \\
	\eq N_{\alpha 6} = \frac{1}{3} \left( u_y B_\alpha - u_\alpha B_y \right) &
	\eq N_{\alpha 7} = \frac{1}{3} \left( u_x B_\alpha - u_\alpha B_x \right) &
	\eq N_{\alpha 8} = \frac{B_\alpha}{9}, \qquad \alpha = x, y
	\end{array}
	\end{equation}
	
	\section{Entropic Method and its Partial Extension to MHD}\label{Sec:Entropic}
	The Karlin group \cite{entropic1,entropic2} introduces the entropic procedure for Navier-Stokes flows by separating the scalar lattice Boltzmann distribution into various moment-related groups. 
	In particular, 
	\begin{equation}\label{fExpansion}
	f_i = k_i + s_i + h_i   \quad , \quad i=0 .. 8
	\end{equation}
	where the $k_i$ distributions correspond to those distributions with conserved moments, the $s_i$ distributions correspond to the stress/shear moments, and finally the $h_i$ distributions correspond to the remaining higher order moments.  Thus for the $k_i$ distributions       
	\begin{equation}\label{contributionExample}
	k_i = \sum_{j=0}^8 \sum_{m=0}^2 \mathrm{T}^{-1}_{im} \mathrm{T}_{mj} f_j   \quad , \quad i=0 .. 8
	\end{equation}
	with the $m-$summation running from $m=0,1,2$ since there are 3 conserved moments.
	Similarly for $s_i$ and $h_i$. 
	The $s_i$ distributions corresponding to the stress/shear moments will come from the set
	\begin{equation}\label{stressContribution}
	s_i \in \left \lbrace d, d \cup t, d \cup q, d \cup t \cup q \right \rbrace
	\end{equation}
	where $d$ is the deviatoric stress, $t$ is the trace of the stress tensor, and $q$ represents the third order moments. Here we choose, for simplicity, the moment contributions $s_i$ to be $d \cup t$ so that 
		\begin{equation}\label{contributionExample1}
	s_i = \sum_{j=0}^8 \sum_{m=3}^5 \mathrm{T}^{-1}_{im} \mathrm{T}_{mj} f_j   \quad , \quad i=0 .. 8  .
	\end{equation}
	Moments 3, 4, and 5 are each second order moments in the $D2Q9$ model, and thus represent the second order quantities $d \cup t$. The moment contributions to $h_i$ are then all the remaining moments that do not contribute to either $k_i$ or $s_i$. Thus 
  \begin{equation}\label{contributionExample2}
	h_i = \sum_{j=0}^8 \sum_{m=6}^8 \mathrm{T}^{-1}_{im} \mathrm{T}_{mj} f_j   \quad , \quad i=0 .. 8  .
	\end{equation}
	
	Karlin et. al. \cite{entropic1,entropic2} now consider the entropy of the post-collisional state, and introduce a parameter $\gamma$ which yields an extremal to this entropy function.  In MRT only some of the relaxation rates affect the transport coefficient under Chapman-Enskog expansions \cite{Pope}.  The transport coefficient in Navier-Stokes simulations is first affected by the stress related distributions $(s_i)$.  The tunable parameter $\gamma$ is introduced to replace the relaxation rates for the higher order moment effects arising from the $(h_i)$ distributions.  In particular, one moves from the standard post-collisional distributions
	\begin{equation}
	f_i^{'} \equiv f_i\left(t+1\right) = f_i + 2\beta \left(\eq f_i - f_i \right)
	\end{equation}
	\begin{equation}\label{postcollisiongamma}
	\text{to} \qquad f_i^{'} = f_i - 2\beta \Delta s_i - \beta \gamma \Delta h_i
	\end{equation}
	where $\beta$ is related to the kinematic viscosity as $\nu = \frac{1}{6} \left( \frac{1}{\beta} - 1 \right)$ and $\Delta s_i = s_i - \eq s_i$, $\Delta h_i = h_i - \eq h_i$, while for the conserved moments $\Delta k_i = k_i - \eq k_i = 0$.
	
	In order to maximize the entropy $S \left[ f \right] $ 
	\begin{equation}
	S \left[ f \right] = - \sum_i f_i \ln(\frac{f_i}{w_i}).
	\end{equation}
	one now writes the entropy in terms of the post-collisional state and the $\gamma$ parameter.  The critical point of the entropy \cite{entropic1,entropic2} determines the tunable parameter $\gamma$ from 
	
	
	\begin{equation}\label{CritPoint}
	\sum_i \Delta h_i \ln(1 + \frac{\left( 1 - \beta \gamma \right) \Delta h_i - \left( 2\beta - 1 \right) \Delta s_i}{\eq f_i}) = 0
	\end{equation}
	This is a rather computationally expensive root-finding procedure having to be done at every point of the grid and at every time step.  Karlin \textit{et. al }\cite{entropic1,entropic2} noted
	that if one invokes the simple small argument expansion  $\log(1+x)=x+...$ one can then readily determine the entropic factor algebraically.  The parameter determined algebraically is denoted by $\gamma^{*}$ :
	\begin{equation}\label{gammaSolved}
	\gamma^{*} = \frac{1}{\beta} - \left( 2 - \frac{1}{\beta} \right) \frac{\left\langle \Delta s | \Delta h \right\rangle }{\left\langle \Delta h | \Delta h \right\rangle }
	\end{equation}
	\begin{equation}
	\text{where the inner product}\quad \left\langle A | B \right\rangle = \sum_i \frac{A_i B_i}{\eq f_i}.
	\end{equation}
	On substituting $\gamma^{*}$ back into the new post-collisional state (Eq. \ref{postcollisiongamma}) a maximal entropy state has been determined for Navier-Stokes flows.  The Karlin group successfully tested this approximation for the tunable parameter $\gamma^{*}(\vec x,t)$ in various simulations of 2D and 3D Navier-Stokes \cite{entropic1,entropic2}.  One thus sees that this emtropic algorithm is a subset of MRT - but it has a dynamic entropic parameter determined at every lattice point and every time step algebraically for entropic stabilization as opposed to the static relaxation times for typical MRT models.
	
	Clearly, this analysis does not simply carry over to LB-MHD with possible non-positive vector magnetic distributions.  Hence we make the ansatz for our partial entropic algorithm that the entropic parameter in LB-MHD is still determined by Eq. (\ref{gammaSolved}) for the corresponding LB-MHD $\Delta h$ and $\Delta s$.  The validity of our ansatz will now be tested against various 2D MHD simulations.
	
	Summarizing, our partial entropic LB-MHD algorithm consists of the following steps (c.f.,  Karlin \textit{et. al.} \cite{entropic1}:
	\begin{enumerate}
		\itemsep 0em
		\item Compute the conserved moments ($\rho$,$\mathbf u$,$\mathbf B$) (Eq. \ref{Transformation}, \ref{KinMomentEq}, \ref{MagMomentEq})
		\item Evaluate the equilibria $\left( f_i^{(\mathrm{eq})} \left( \rho, \mathbf u, \mathbf B \right), \vec g_k^{(\mathrm{eq})} \left( \rho, \mathbf u, \mathbf B \right) \right)$ (Eq. \ref{feq})
		\item Compute $s$ and $s^{(\mathrm{eq})}$ (Eq. \ref{contributionExample}, \ref{stressContribution})
		\item Compute $\Delta s_i = s_i - \eq s_i$
		\item Compute $\Delta h_i = h_i - \eq h_i = f_i - \eq f_i - \Delta s_i$
		\item Evaluate $\gamma^{*}$ (Eq. \ref{gammaSolved})
		\item Relax (Collide): $f_i^{'}$ (Eq. \ref{postcollisiongamma}),  and corresponding $\vec g_k^{'}$.
	\end{enumerate}
	
	Standard LB-MHD is recovered for entropy parameter:   $\gamma(\vec x,t) = const. = 2$.  As mentioned earlier, there is no attempt made to find a corresponding maximal entropy state for the magnetic distribution function since the magnetic field in most problems of interest undergoes field reversal.  (e.g., in magnetic field reconnection..).  However the effect of working with the maximal entropy state for the particle distribution function will have direct effects on the evolution of the magnetic field due to the coupling of the $\vec B$-field in the relaxation distribution function $\eq f$ as well as the coupling of the fluid velocity $\vec u$ in $\eq{\vec g}$.
	
	
	\section{Partially Entropic LB-MHD Simulations}\label{Sec:Simulation}
	We first have benchmarked our partially entropic LB-MHD code against our earlier (totally non-entropic) MRT LB-MHD simulations of  a Kelvin-Helmholtz jet instability in a magnetic field \cite{flint9bit}.  Here we show the physics recovered by the variations in the partially entropic parameter $\gamma^{*}$  and its variations away from the MRT value of $\gamma^{*}(\vec x,t) \equiv 2.0$ for sufficiently weak axial $\vec B$ that the jet is unstable.  Some runs were then performed to examine the increased numerical stability in the parameter regime of the mean velocity $\vec u$ and magnetic field $\vec B$ due to the partially entropic algorithm.  Following this we consider the interplay between Kelvin-Helmholtz instability and the tearing mode instability and qualitatively compare our results to that of Chen et. al,  \cite{Chen}.  Finally we qualitatively compare our simulations with the Biskamp-Welter profile.

	\subsection{Magnetized Kelvin-Hemholtz Jet Instability}
	
\begin{figure}[!htb]
		\centering
		\begin{subfigure}{0.45\textwidth}
			\centering
			\includegraphics[height=1.59in, keepaspectratio=true]{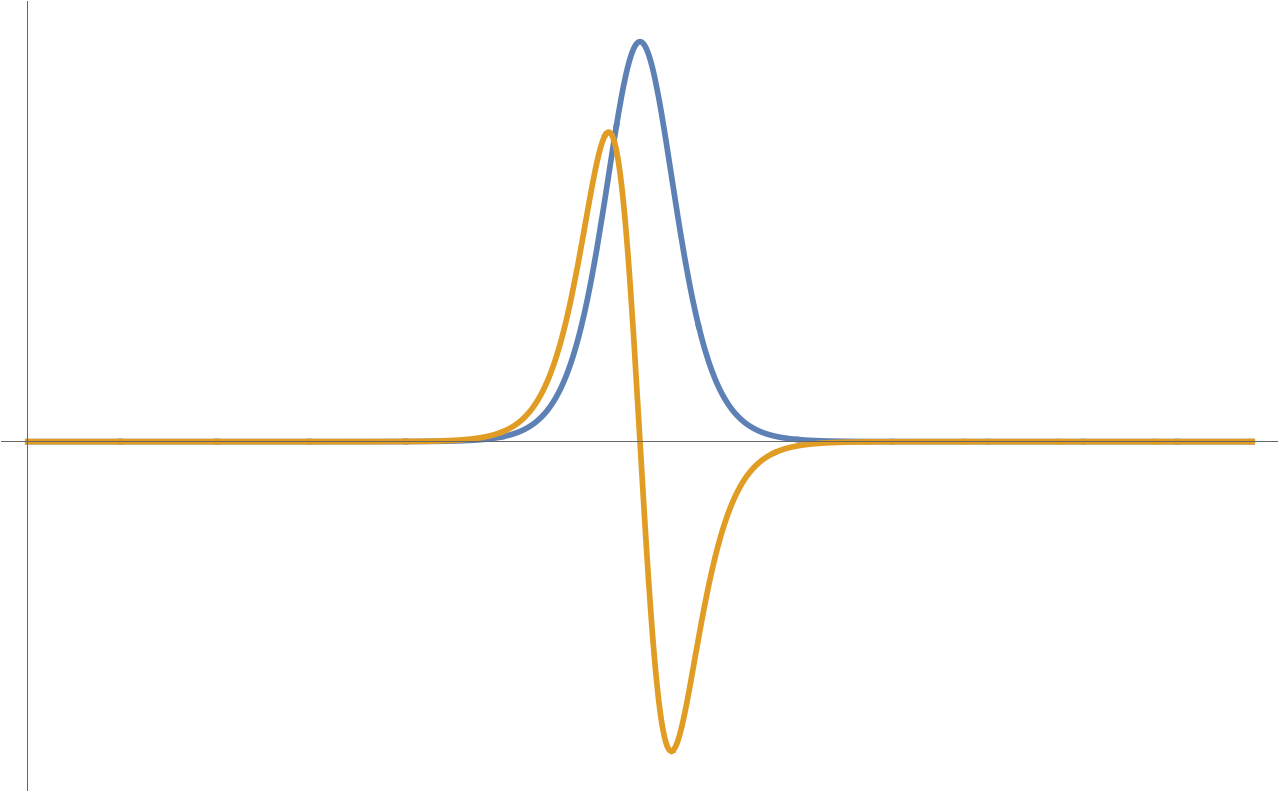}
			\caption{}
		\end{subfigure}
		\begin{subfigure}{0.45\textwidth}
			\centering
			\includegraphics[height=1.59in, keepaspectratio=true]{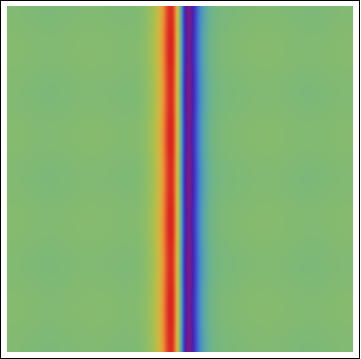}
			\caption{}
		\end{subfigure}
		\caption{\label{KHInitProf}(a) The Initial velocity (blue) and vorticity (orange) in the unstable magnetized jet simulation as a function of $x$. (b) The corresponding initial vorticity $\omega(x,y)$: red for $\omega > 0$ and blue for $\omega < 0$}
		
	\end{figure}

	We now consider the partially entropic-LB-MHD algorithm for the breakdown of a Kelvin-Helmholtz jet in a weak magnetic field.  In our simulations, the initial parameters are so chosen that there is a direct cascade of energy to small spatial scales (indicating the existence of a magnetic field) but the magnetic field is sufficient weak so as not to stabilize the jet, Fig. \ref{KHInitProf} 
	
	\begin{equation}
	\vec u(t=0) = U_0 \sech^2 (x) \hat y  ,\quad  \vec B(t=0) = B_0 \hat y
	\end{equation}
	
\begin{figure}[!hp]
		\centering
		\begin{tabular}{rccc}
			$t=44$k&	
			\raisebox{-0.5\height}{\includegraphics[height=1.25in, keepaspectratio=true]{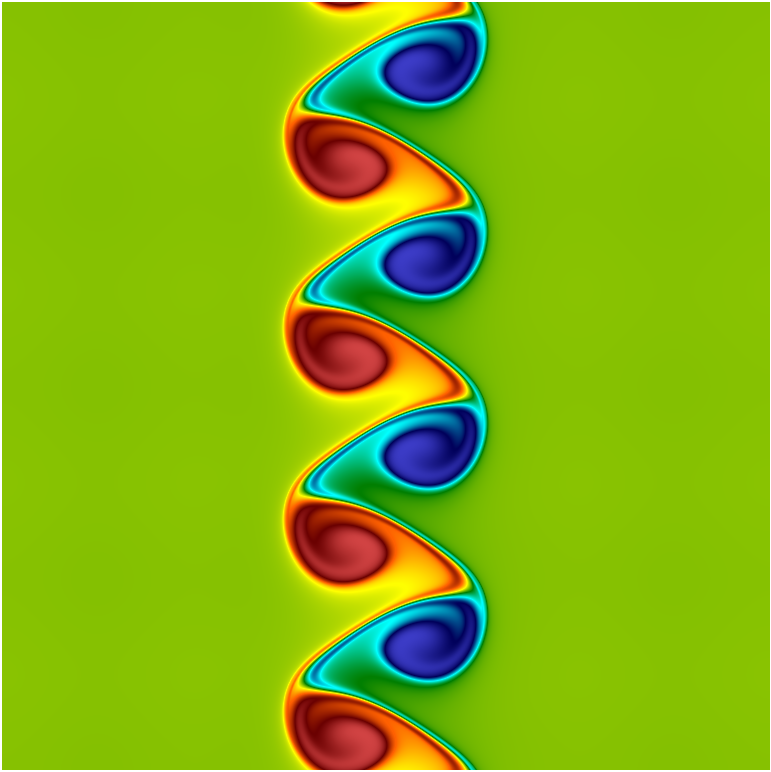}}&
			\raisebox{-0.5\height}{\includegraphics[height=1.25in, keepaspectratio=true]{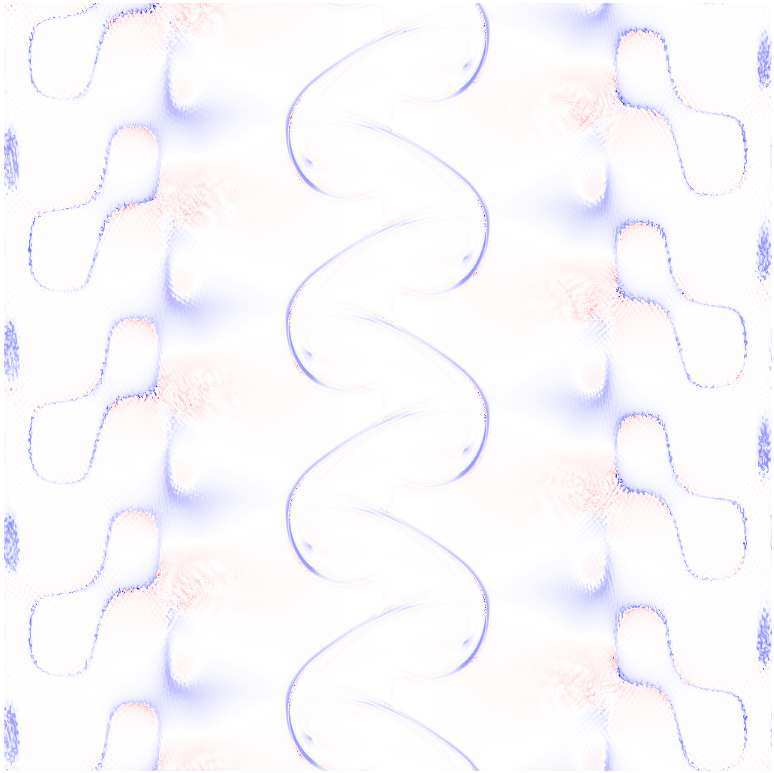}}&
			\raisebox{-0.5\height}{\includegraphics[height=1.25in, keepaspectratio=true]{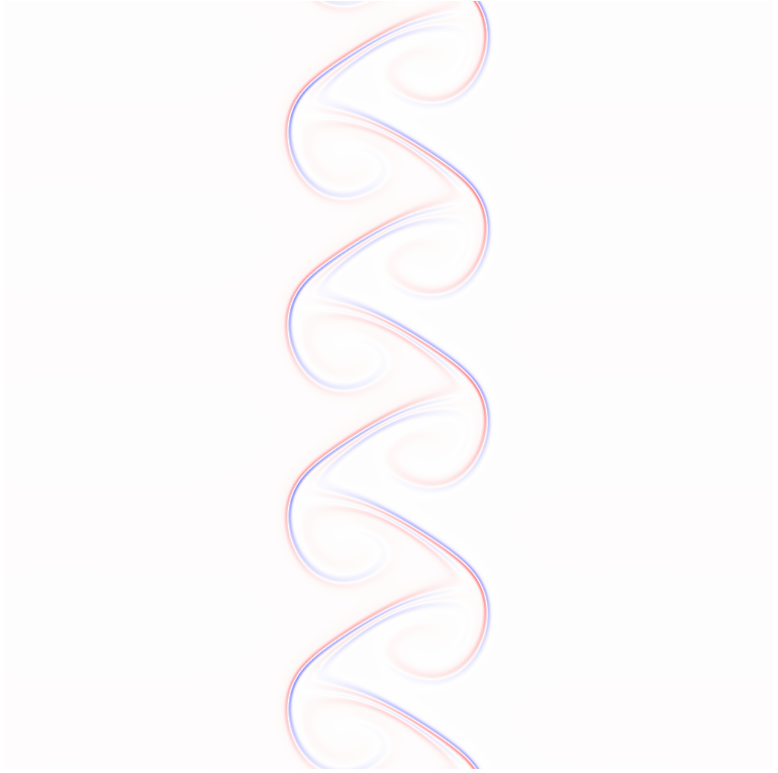}}\\
			$t=80$k&	
			\raisebox{-0.5\height}{\includegraphics[height=1.25in, keepaspectratio=true]{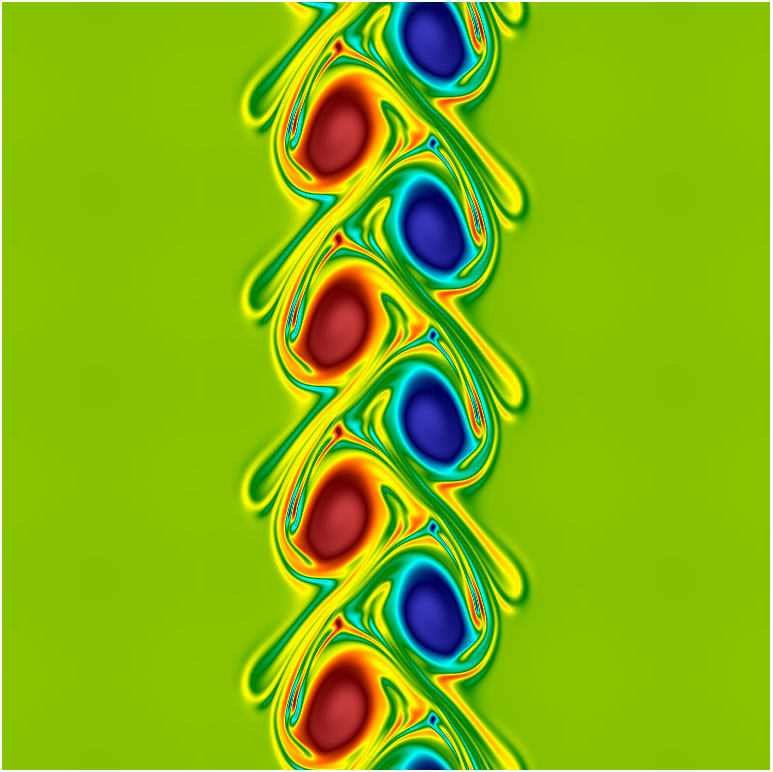}}&
			\raisebox{-0.5\height}{\includegraphics[height=1.25in, keepaspectratio=true]{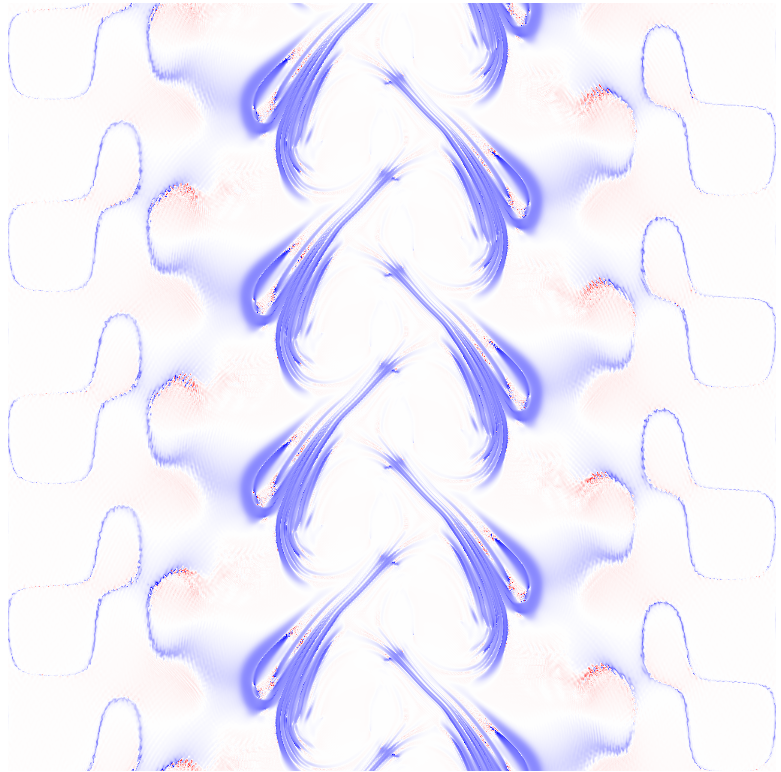}}&
			\raisebox{-0.5\height}{\includegraphics[height=1.25in, keepaspectratio=true]{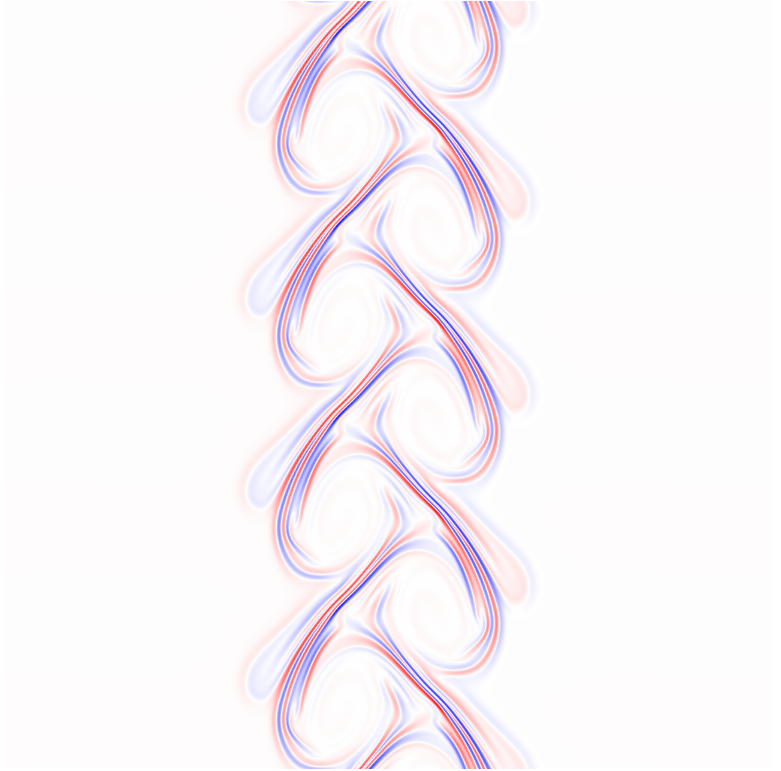}}\\
			$t=266$k&	
			\raisebox{-0.5\height}{\includegraphics[height=1.25in, keepaspectratio=true]{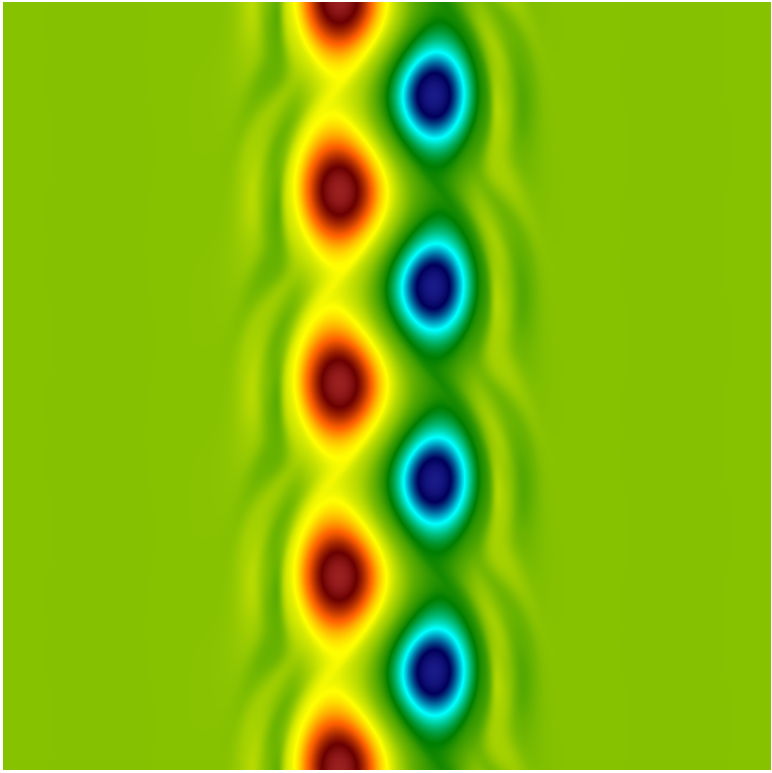}}&
			\raisebox{-0.5\height}{\includegraphics[height=1.25in, keepaspectratio=true]{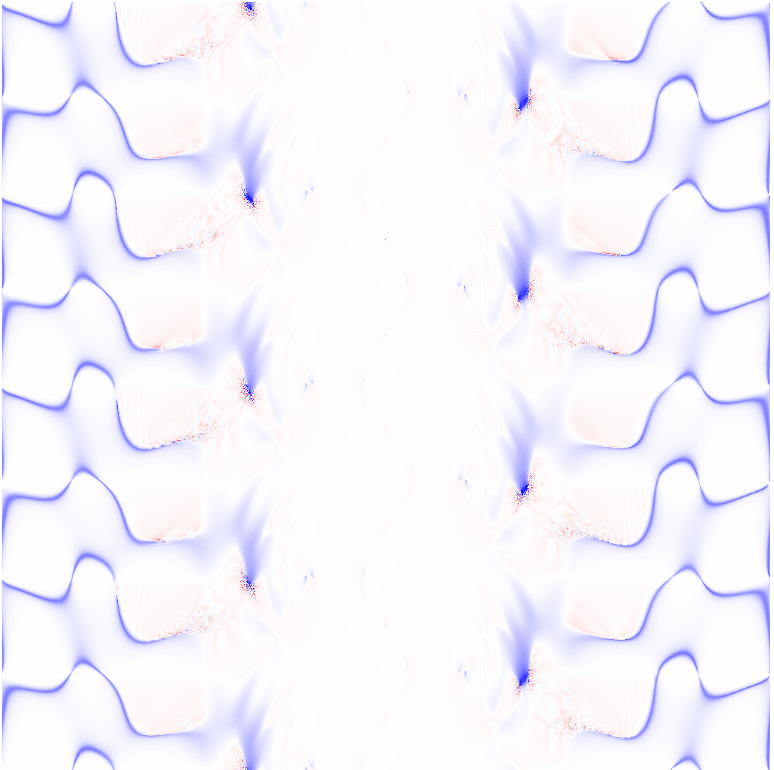}}&
			\raisebox{-0.5\height}{\includegraphics[height=1.25in, keepaspectratio=true]{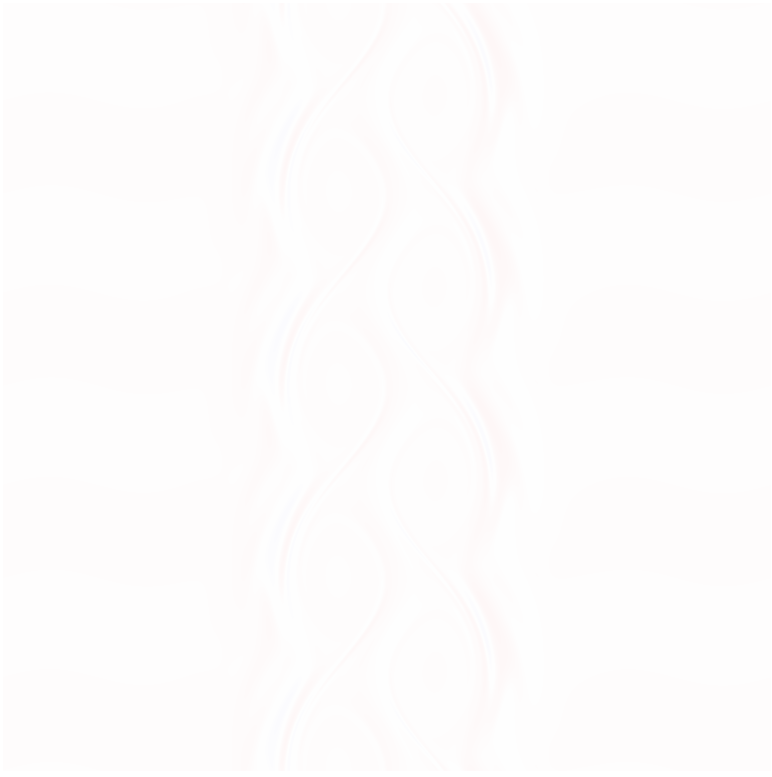}}\\
			$t=344$k&	
			\raisebox{-0.5\height}{\includegraphics[height=1.25in, keepaspectratio=true]{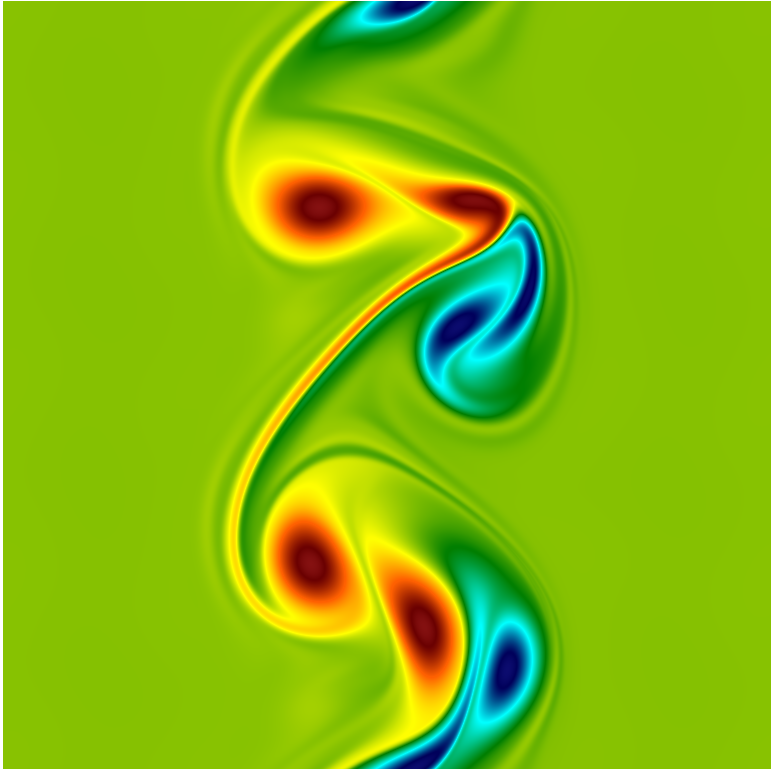}}&
			\raisebox{-0.5\height}{\includegraphics[height=1.25in, keepaspectratio=true]{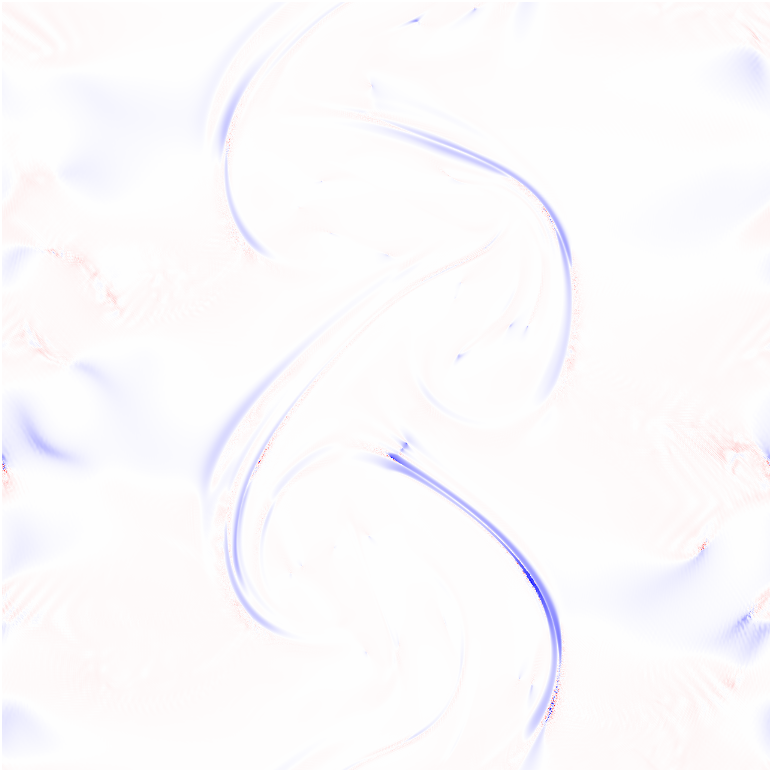}}&
			\raisebox{-0.5\height}{\includegraphics[height=1.25in, keepaspectratio=true]{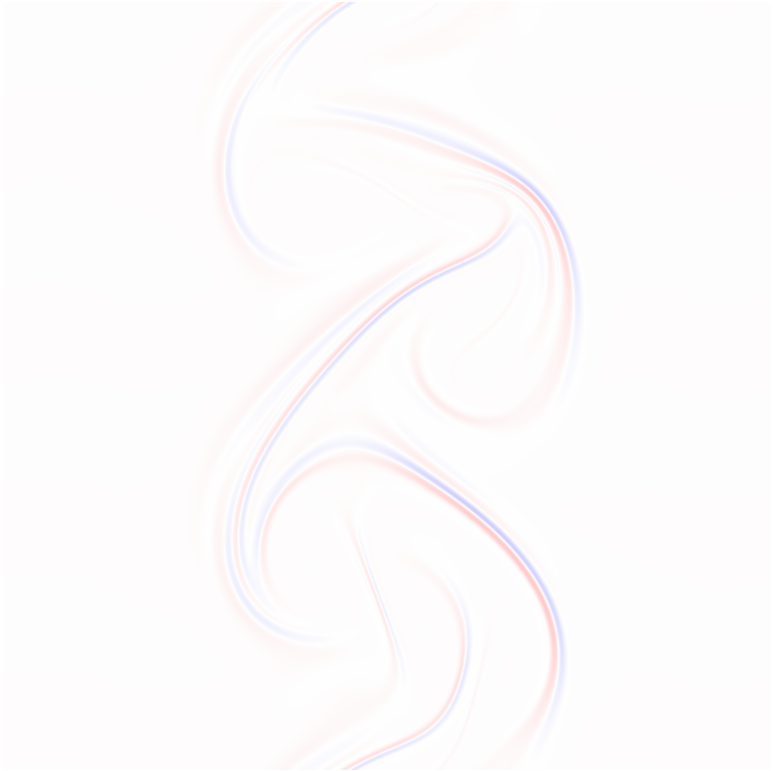}}\\
			$t=500$k&	
			\raisebox{-0.5\height}{\includegraphics[height=1.25in, keepaspectratio=true]{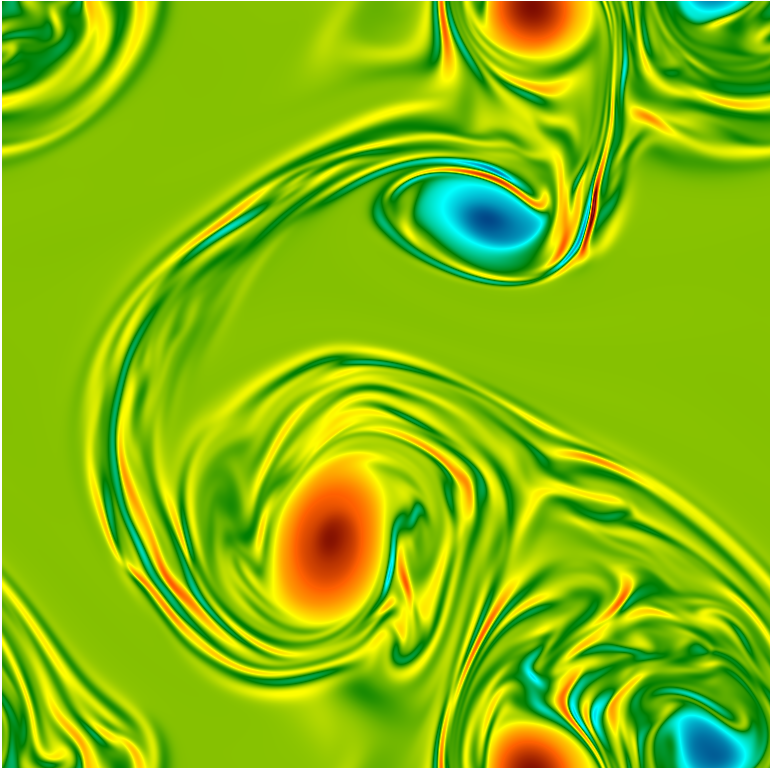}}&
			\raisebox{-0.5\height}{\includegraphics[height=1.25in, keepaspectratio=true]{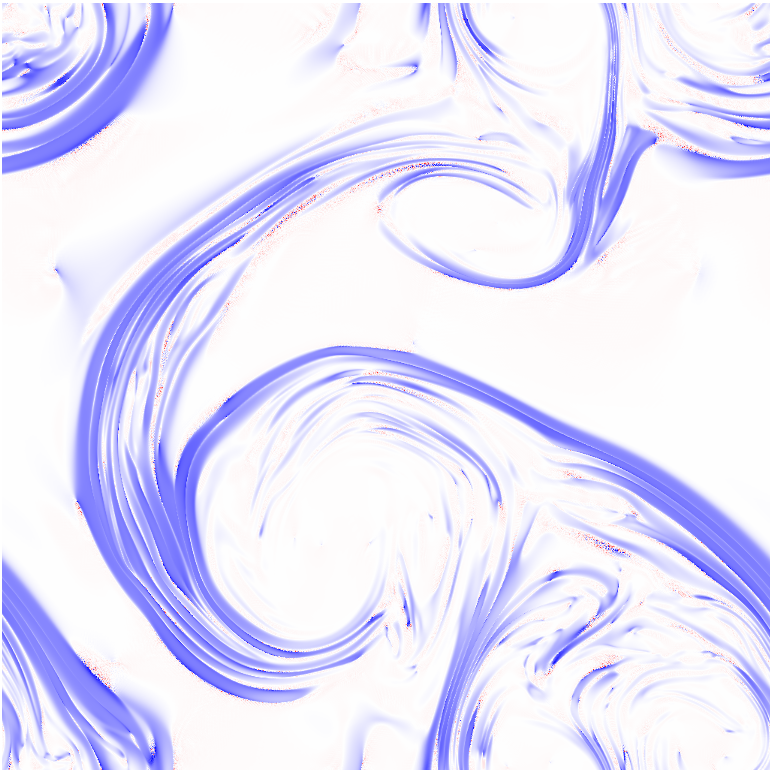}}&
			\raisebox{-0.5\height}{\includegraphics[height=1.25in, keepaspectratio=true]{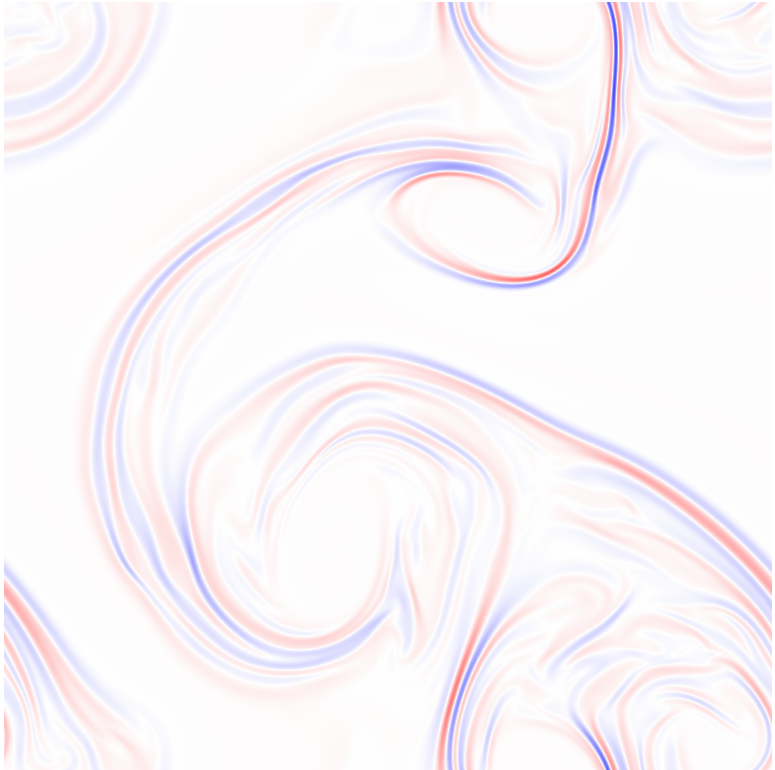}} \vspace*{2mm} \\&
			\raisebox{-0.5\height}{\includegraphics[height=.8in, keepaspectratio=true]{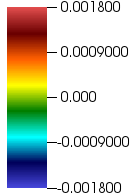}}&
			\raisebox{-0.5\height}{\includegraphics[height=.8in, keepaspectratio=true]{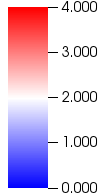}}&
			\raisebox{-0.5\height}{\includegraphics[height=.8in, keepaspectratio=true]{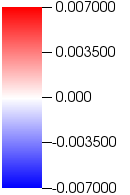}}\\
			\\
			& (a) & (b) & (c)
		\end{tabular}
		\caption[.]{\label{LowBw}Evolution of a Kelvin-Helmholtz jet with very weak axial magnetic field:   $B_0=0.005U_0$. The column 2D plots are for (a) the vorticity $\omega$, (b) the entropy parameter $\gamma^*$, and (c) the current $j$.  The jet is unstable forming a von-Karman like vortex street (time $t = 44k$).  These vortices start to generate secondary smaller vortex streaks ($t = 80k$) - where the entropy factor becomes important.  The vortex street then becomes unstable $t=344k$) with vortex-vortex reconnection dominating shortly after the break-up of the vortex street.  However by $t=500k$ strong subsidiary vortices are generated because of the 2D MHD turbulence with significant corresponding regions of variations of the entropic parameter away from $2$.   Note that the color scheme is held constant for all time snapshots.  Spatial grid $1024^2$}  
		
	\end{figure}

\begin{figure}[!htb]
	\centering
	\includegraphics[height=1.5in, keepaspectratio=true]{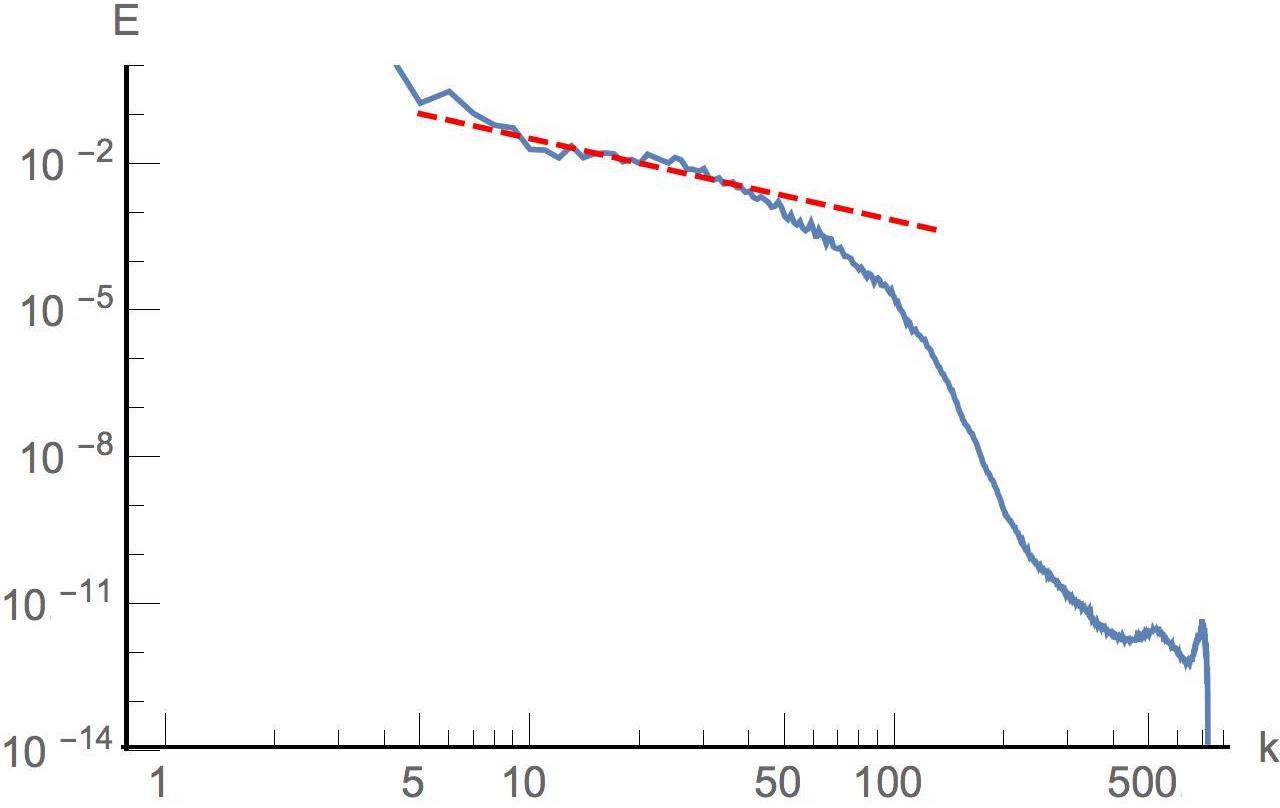}
	\caption[.]{\label{KHtotEn500k}Spectral plot of the Kelvin-Helmholtz simulation at $t=500k$ where gridsize is $1024^2$ and the slope of the dashed line is $k^{-1.67}$.}
	
	\end{figure}

	The evolution of the vorticity, $\omega$, the current, $j$ as well as the entropic stabilization parameter $\gamma^*$ for this 2D jet is plotted in Fig. \ref{LowBw}.  With the (dimensionless) choice of $B_0=0.005 U_0$, the jet breaks into a Kelvin-Helmholtz vortex street ($t \leq 266k$).  There is then further symmetry breaking as the vortex street is broken up, leading to vortex-vortex reconnection (ala 2D Navier-Stokes turbulence), as well as
	the generation of small scales eddies (characteristic of 2D MHD) for $t > 266k$.
	One notices that the partial entropy parameter $\gamma^{*}$ in fig. \ref{LowBw} deviates from the ordinary LB-MHD value of $\gamma^{*}(\vec x,t) = 2$ wherever there are significant number of small eddies.  These are regions of steep gradients and it is in these regions where the partial entropic stabilization of the simulation occurs.  It is important to note that this partial entropy stabilization is occurring from local information at each lattice site.  This is reminiscent of LB where gradients can be computed from local moments of perturbed distributions:  e.g., in large eddy simulation modelings in the Smagorinsky model, the mean velocity gradients are determined from simple local moments. 
	For stronger magnetic fields, the jet will be stabilized and is of little interest for our partial entropic-LB-MHD model, \cite{flint9bit}.
	A spectral plot of the total energy of the Kelvin-Helmholtz simulation at $t=500$k is presented in fig. \ref{KHtotEn500k} with a slope of $k^{-\frac{5}{3}}$. This spectral plot corresponds to the final timestep in Fig. \ref{LowBw}.
	
	\subsubsection{Stability Improvements from the Entropic Algorithm}
	Some numerical stability boundaries were investigated between ordinary LB-MHD and our partial entropic-LB-MHD with the $\gamma^*$ parameter on a grid of $1024^2$ for the Kelvin-Helmholtz jet.  We found that the partial entropic-LB-MHD algorithm permitted a maximal mean velocity $\vec U_{0,max}$ to be increased by a factor of $2$ in a purely Navier-Stokes turbulence simulation (i.e., no $\vec B$-field) while the velocity maximum could be increased by a factor of $8$ when there was a strong stabilizing $\vec B$-field.  As regards the magnetic field (at fixed $\vec U_0$), the partial entropic-LB-MHD algorithm permitted an increase by a factor of $2$ in $\vec B_0$.  In the partial entropic-LB-MHD algorithm the viscosity could become arbitrary small, while ordinary LB-MHD the minimum stable viscosity was $10^{-5}$ when $\vec B_0 = 0$, and $10^{-2}$ when there was a strong stabilizing $\vec B_0$.  No substantial stability limits were found on the achievable minimum resistivity.
	
	It should be stressed that the computational overhead of computing this entropic parameter $\gamma^*$ is quite small, primarily because it is determined algebraically from local information only.
	
	
		
	\begin{figure}[!htb]
	\centering
	\begin{subfigure}[b]{0.15\textwidth}
		\centering
		\includegraphics[width=0.75in, height=2.54in, keepaspectratio=true]{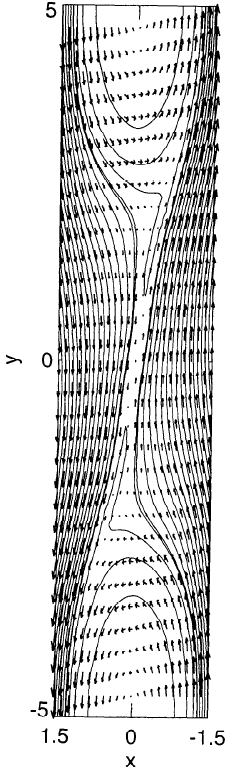}
		\caption{}
	\end{subfigure}
	\begin{subfigure}[b]{0.15\textwidth}
		\centering
		\includegraphics[width=0.7in, height=2.35in, keepaspectratio=true]{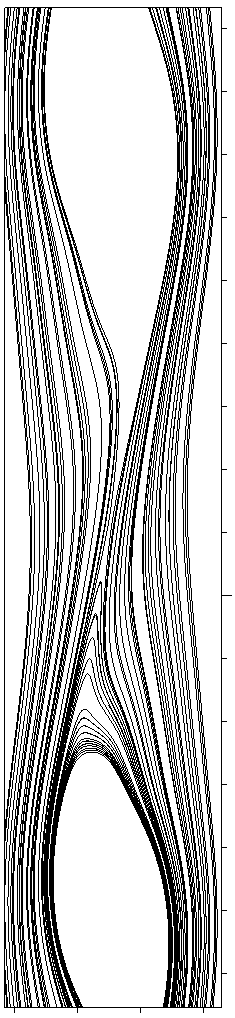}
		\vspace*{4mm}
		\caption{}
	\end{subfigure}
	\caption{\label{ChenCase13Orig} A snapshot of the 2D spatial dependence of (a) magnetic field lines (and velocity fields) from a Chen et. al. supersonic Alfvenic simulation, their Fig. 6a, compared to (b) our entropic LB-MHD simulation on $1024^2$ grid for the same initial profiles. $S = 1000.$}
	\end{figure}

	\begin{figure}[!htb]
	\centering
	\begin{subfigure}[b]{0.15\textwidth}
		\centering
		\includegraphics[width=0.77in, height=2.59in, keepaspectratio=true]{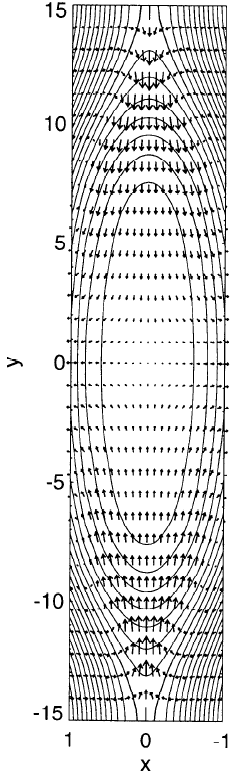}
		\caption{}
	\end{subfigure}
	\begin{subfigure}[b]{0.15\textwidth}
		\centering
		\includegraphics[width=0.7in, height=2.3in, keepaspectratio=true]{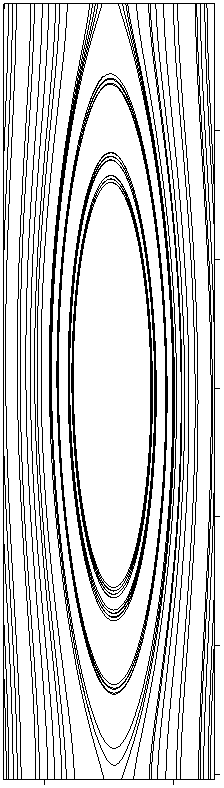}
		\vspace*{4mm}
		\caption{}
	\end{subfigure}
	\caption{\label{ChenCase5Orig}A snapshot of the 2D spatial dependence of the (a) the magnetic field line contours (and velocity field) for zero initial shear ($V_0 = 0.0$), from Chen et. al. Fig. 4a, and (b) from our entropic LB-MHD algorithm on a $1024^2$ grid.}
	\end{figure}

	\begin{figure}[!htb]
	\centering
	\begin{subfigure}[b]{0.15\textwidth}
		\centering
		\includegraphics[width=0.9in, height=2.7in, keepaspectratio=true]{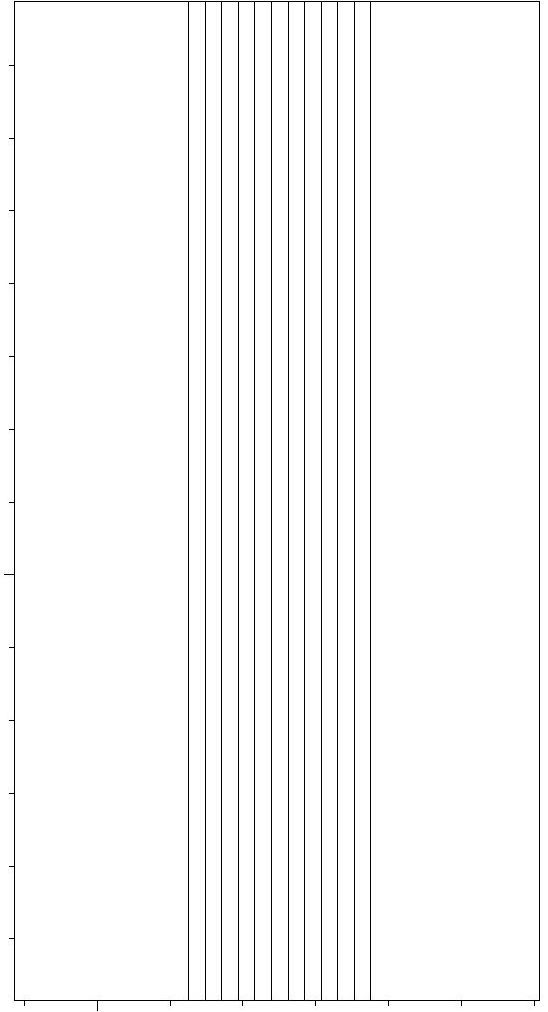}
		\caption{}
	\end{subfigure}
	\begin{subfigure}[b]{0.15\textwidth}
		\centering
		\includegraphics[width=0.9in, height=2.7in, keepaspectratio=true]{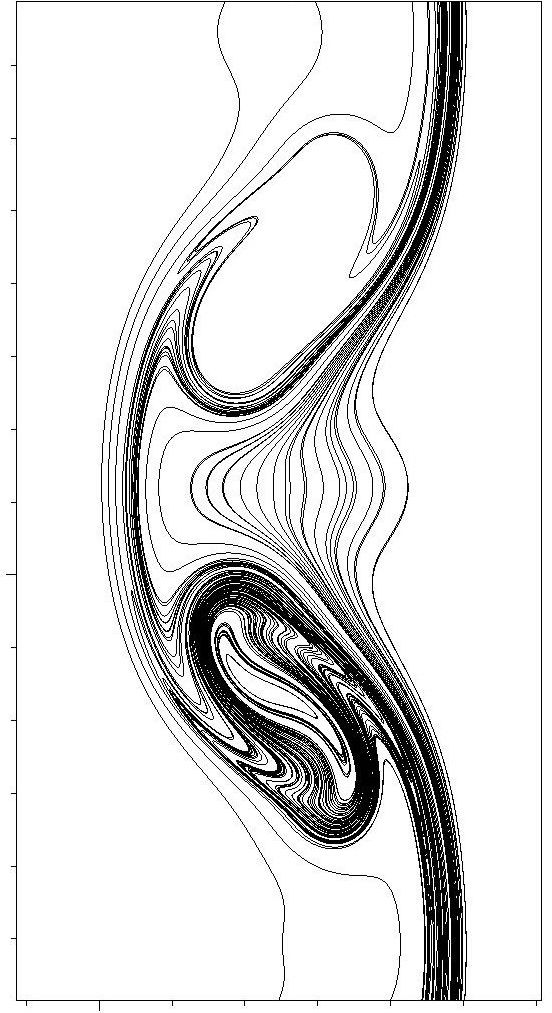}
		\caption{}
	\end{subfigure}
	\begin{subfigure}[b]{0.15\textwidth}
		\centering
		\includegraphics[width=0.9in, height=2.7in, keepaspectratio=true]{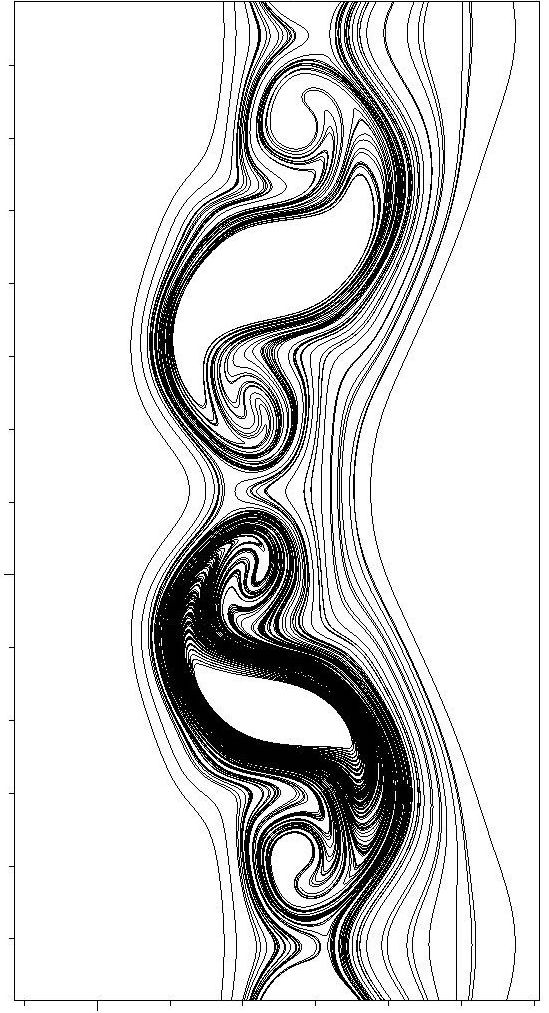}
		\caption{}
	\end{subfigure}
	\begin{subfigure}[b]{0.15\textwidth}
		\centering
		\includegraphics[width=0.9in, height=2.7in, keepaspectratio=true]{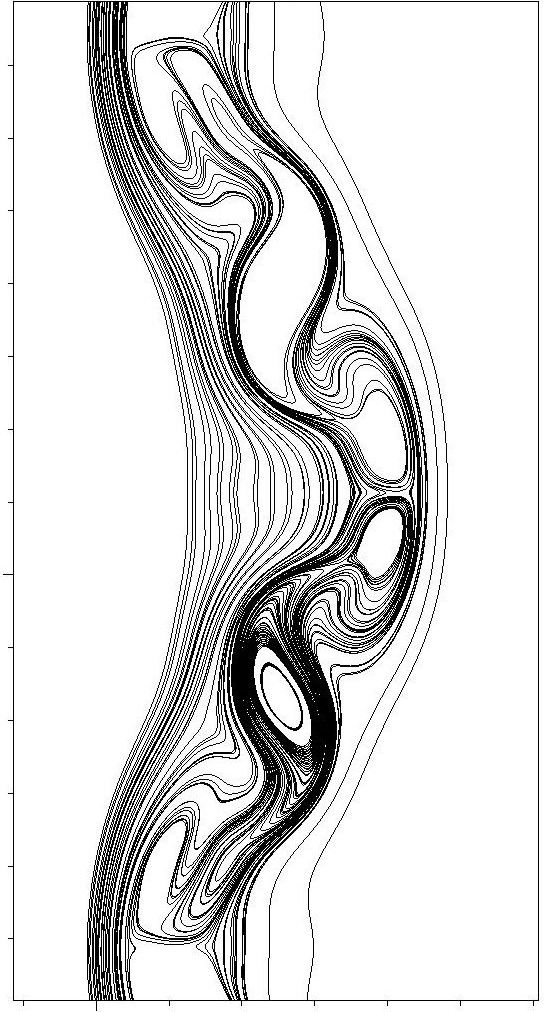}
		\caption{}
	\end{subfigure}
	\begin{subfigure}[b]{0.15\textwidth}
		\centering
		\includegraphics[width=0.9in, height=2.7in, keepaspectratio=true]{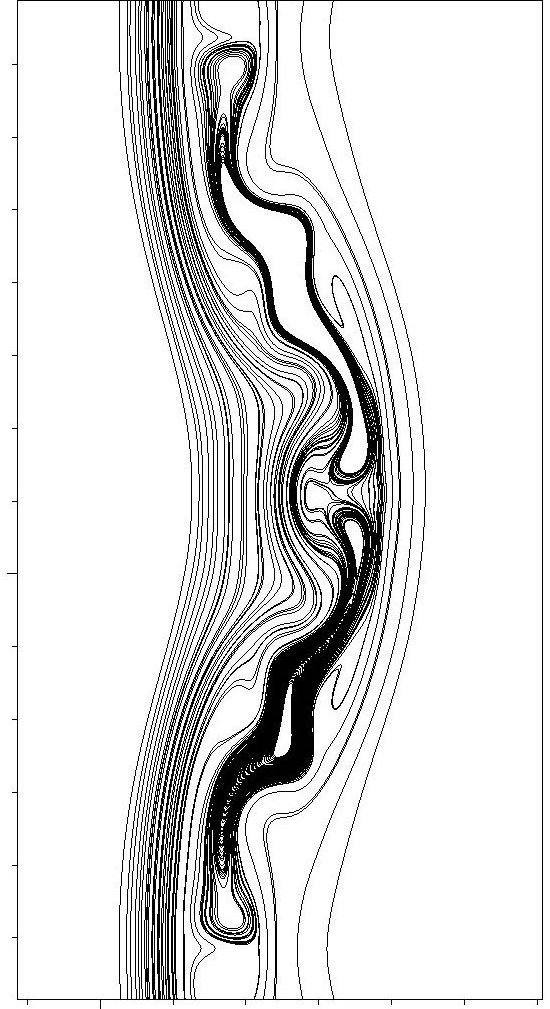}
		\caption{}
	\end{subfigure}
	\begin{subfigure}[b]{0.15\textwidth}
		\centering
		\includegraphics[width=0.9in, height=2.7in, keepaspectratio=true]{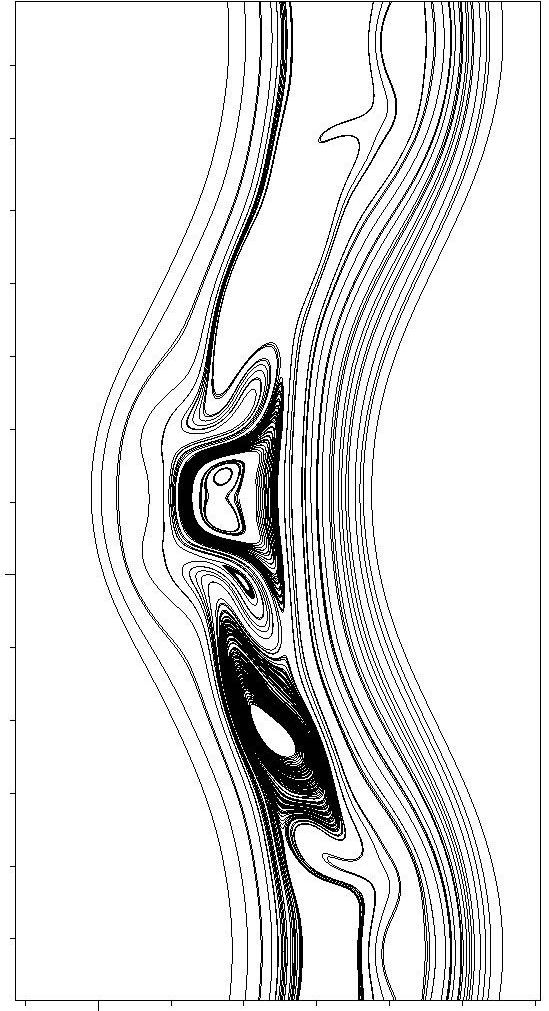}
		\caption{}
	\end{subfigure}
\\
	\begin{subfigure}[b]{0.15\textwidth}
		\centering
		\includegraphics[width=0.9in, height=2.7in, keepaspectratio=true]{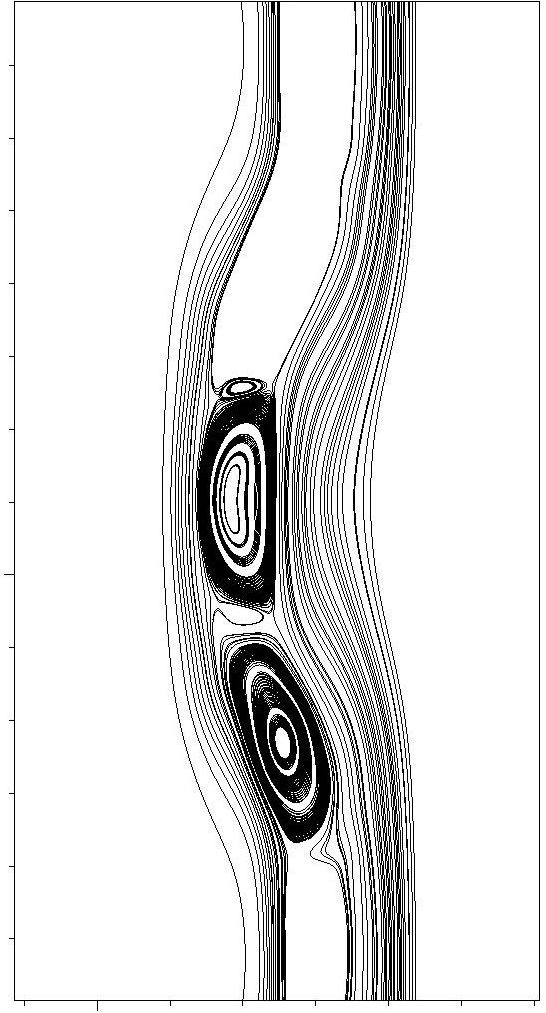}
		\caption{}
	\end{subfigure}
	\begin{subfigure}[b]{0.15\textwidth}
		\centering
		\includegraphics[width=0.9in, height=2.7in, keepaspectratio=true]{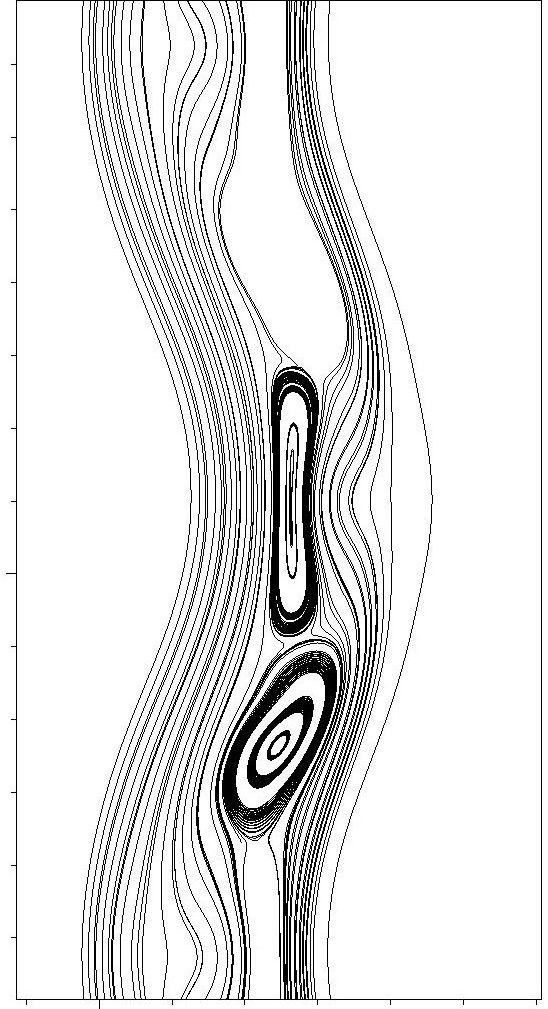}
		\caption{}
	\end{subfigure}
	\begin{subfigure}[b]{0.15\textwidth}
		\centering
		\includegraphics[width=0.9in, height=2.7in, keepaspectratio=true]{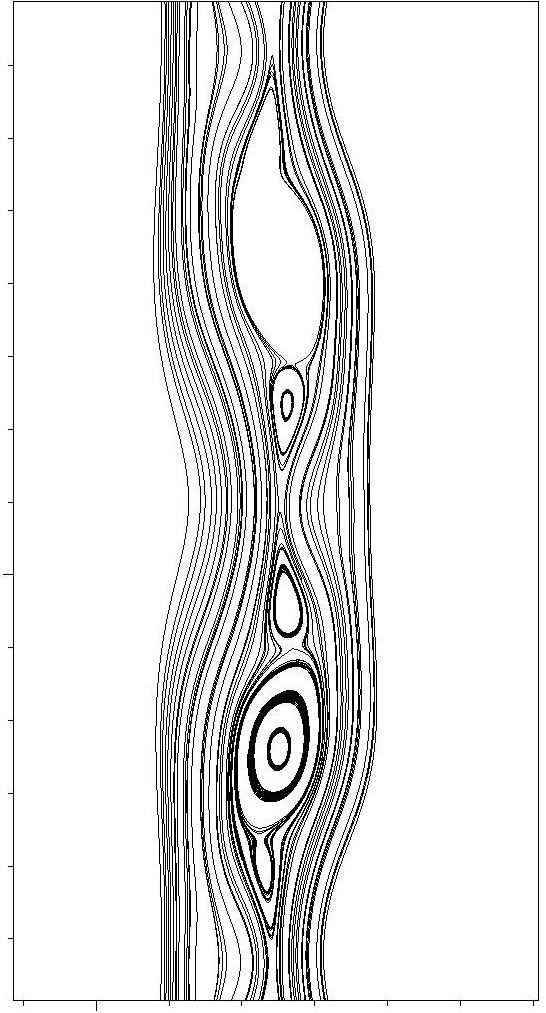}
		\caption{}
	\end{subfigure}
	\begin{subfigure}[b]{0.15\textwidth}
		\centering
		\includegraphics[width=0.9in, height=2.7in, keepaspectratio=true]{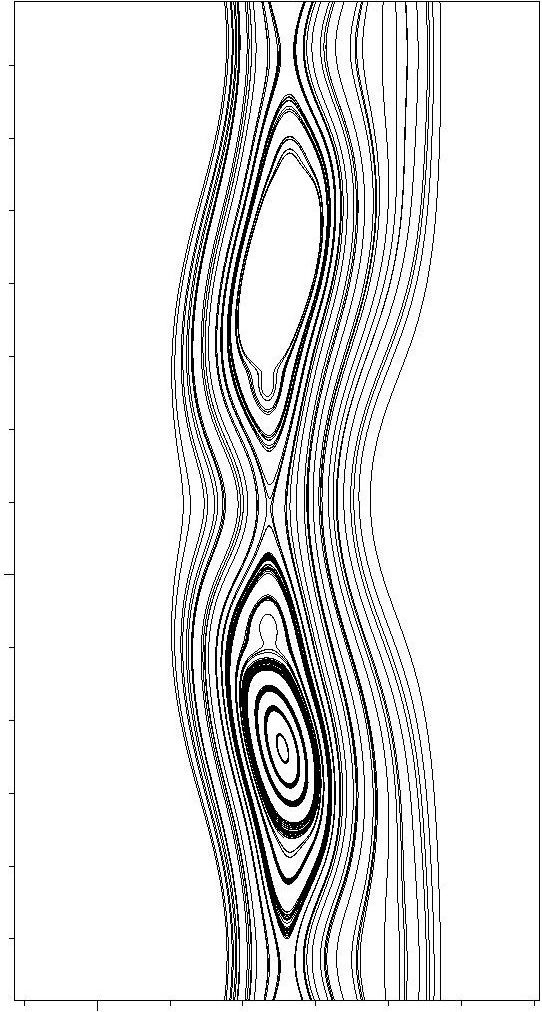}
		\caption{}
	\end{subfigure}
	\begin{subfigure}[b]{0.15\textwidth}
		\centering
		\includegraphics[width=0.9in, height=2.7in, keepaspectratio=true]{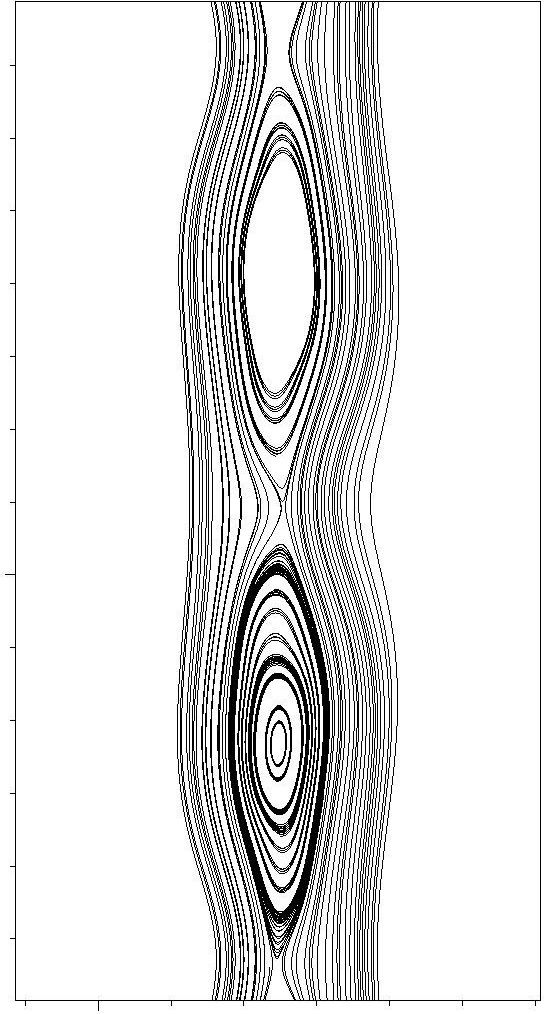}
		\caption{}
	\end{subfigure}
	\begin{subfigure}[b]{0.15\textwidth}
		\centering
		\includegraphics[width=0.9in, height=2.7in, keepaspectratio=true]{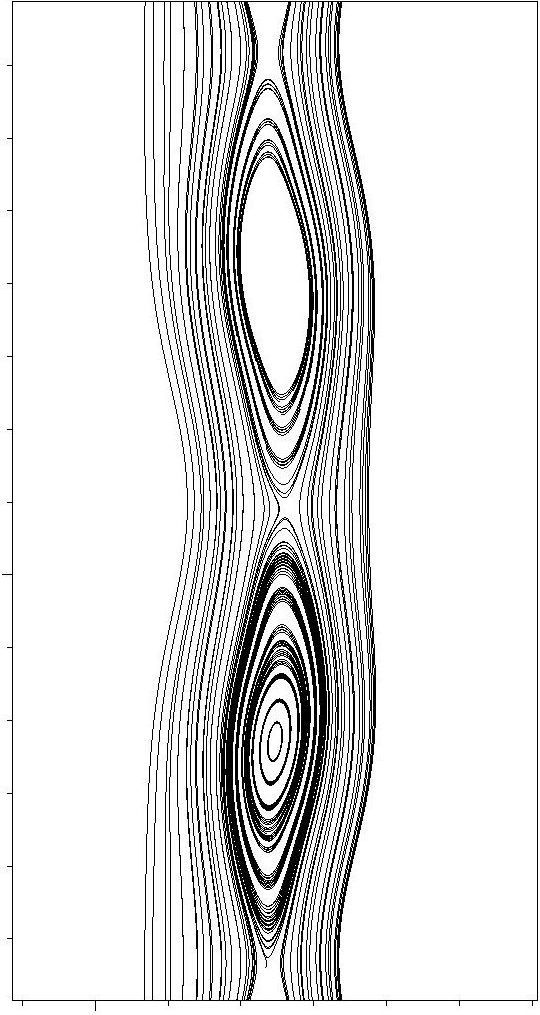}
		\caption{}
	\end{subfigure}
	\caption{\label{ChenCase5Evol}Evolution of the magnetic field lines from our partial entropic LB-MHD code for zero initial shear velocity in a uniform magnetic field.  Snapshots of the field lines are presented at each 8000 (8k) LB time steps.  Grid $1024^2$.  (a) t = 0,  (b) t = 8000, (c) t = 16000, (d) t = 24000, (e) t = 32000, (f) t = 40000, (g) t = 48000, (h) = 56000, (i) 64000, (j) t = 72000, (k) t = 80000, (l) t = 88000.}
	
	\end{figure}

	\subsection{Chen \textit{et. al.} Profile}
	Chen \textit{et. al.} \cite{Chen} has considered the linear and nonlinear evolution of Kelvin-Helmholtz (velocity shear) vs. the tearing mode (magnetic shear) instabilities in 2D compressible MHD.  Their closure includes an evolution equation for the enthalphy as well as various resistivity profiles using standard CFD techniques.  Their initial profiles are
	\begin{equation}
	u_y(x,t=0) = -U_0 \tanh(x)  , \quad
	B_y(x,t=0) = B_0 \tanh(x).
	\end{equation}
	Thus our comparisons can only be qualitative, and we only consider the Chen et. al.  \cite{Chen} simulations when they keep their resistivity constant. 
	Typically, when the velocity is below the Alfven speed, it stabilizes the tearing mode and so reduces the reconnection rate.  However, if the velocity is above the Alfven speed the Kelvin-Helmholtz instability sets in.  In our first partial entropic LB-MHD simulation, we consider super-Alfven velocity shear flow and the Kelvin-Helmholtz induced magnetic islands due to reconnection in Fig. \ref{ChenCase13Orig}.  In Fig. \ref{ChenCase13Orig}a we show the simulation results of case 13 in Chen et. al. for the magnetic field lines and compare them to those arising from our partial entropic LB-MHD model for resistivity $\eta = 0.001$, Fig. \ref{ChenCase13Orig}b.
	
	For the case of no initial shear, large magnetic islands are formed. A corresponding snapshot is given of the magnetic field lines from the case 5 simulation in Chen et. al., Fig. \ref{ChenCase5Orig}a, and from our entropic LB-MHD model, Fig. \ref{ChenCase5Orig}b.
	In Fig. \ref{ChenCase5Evol} we show the partial entropic LB-MHD evolution of the magnetic field lines for this initial zero velocity shear flow parameter set of Fig. \ref{ChenCase5Orig}.   It seems for the case considered here, the enthalpy equation in Chen \cite{Chen} does not play a significant role.

	\subsection{Biskamp-Welter Profile}
\begin{figure}[!htb]
	\centering
	\raisebox{0.7\height}{\includegraphics[height=1.25in, keepaspectratio=true]{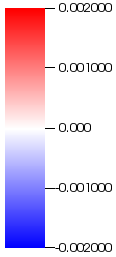}}
	\begin{subfigure}[b]{0.4\textwidth}
		\centering
		\includegraphics[width=3.25in, height=2.0in, keepaspectratio=true]{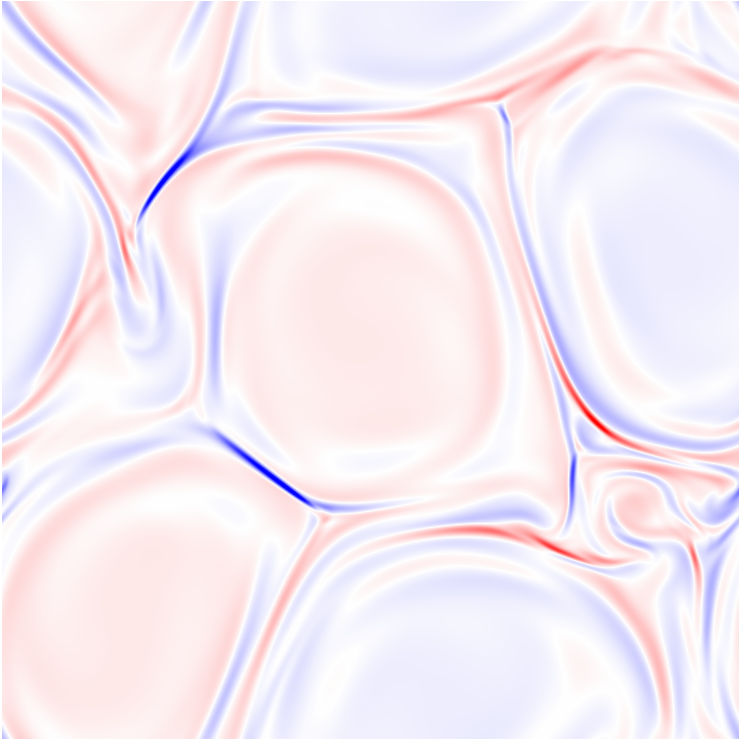}
		\caption{}
	\end{subfigure}
	\begin{subfigure}[b]{0.4\textwidth}
		\centering
		\includegraphics[width=2.59in, height=2.0in, keepaspectratio=true]{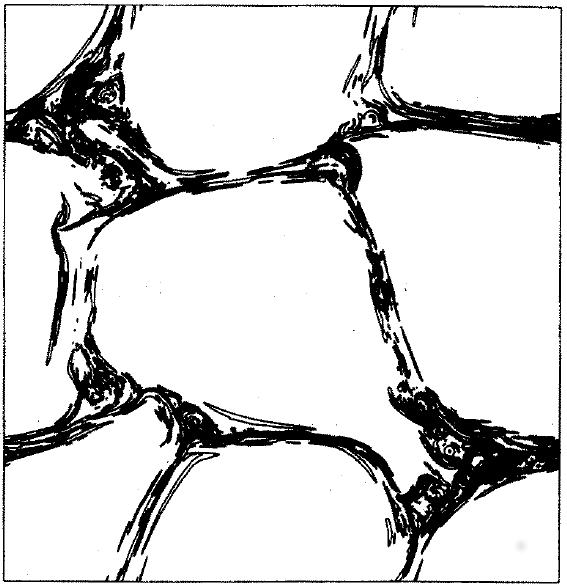}
		\caption{}
	\end{subfigure}
	\caption{\label{BiskampJ} Snapshot of the current lines from (a) partial entropic LB-MHD code on a grid of $1024^2$ at time = 226k, (b) Biskamp-Welter, Fig. 11a}	
	\end{figure}

\begin{figure}[!htb]
	\centering
	\raisebox{0.55\height}{\includegraphics[height=1.25in, keepaspectratio=true]{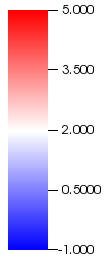}} \quad
	\includegraphics[width=3.25in, height=2.0in, keepaspectratio=true]{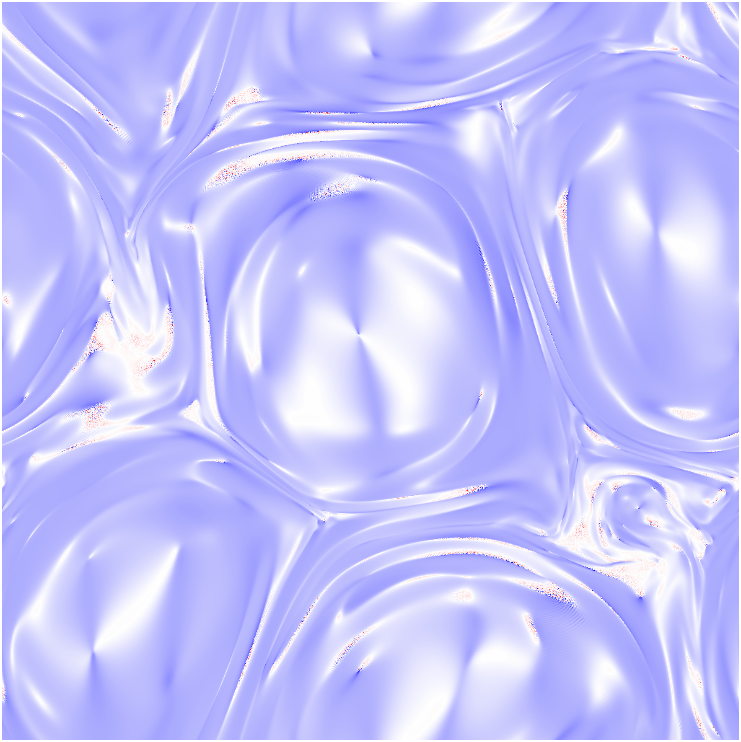}
	\caption[.]{\label{GammaPlot}Plot of the entropic parameter $\gamma^*$ after 226k timesteps on a  $1024^2$ grid.  $\gamma = 2.0$ corresponds to ordinary LB-MHD.  Lattice points with $\gamma^* \neq 2.0$ correspond to the effects of the partial entropic LB-MHD algorithm.}
	\end{figure}

\begin{figure}[!htb]
	\centering
	\includegraphics[height=1.5in, keepaspectratio=true]{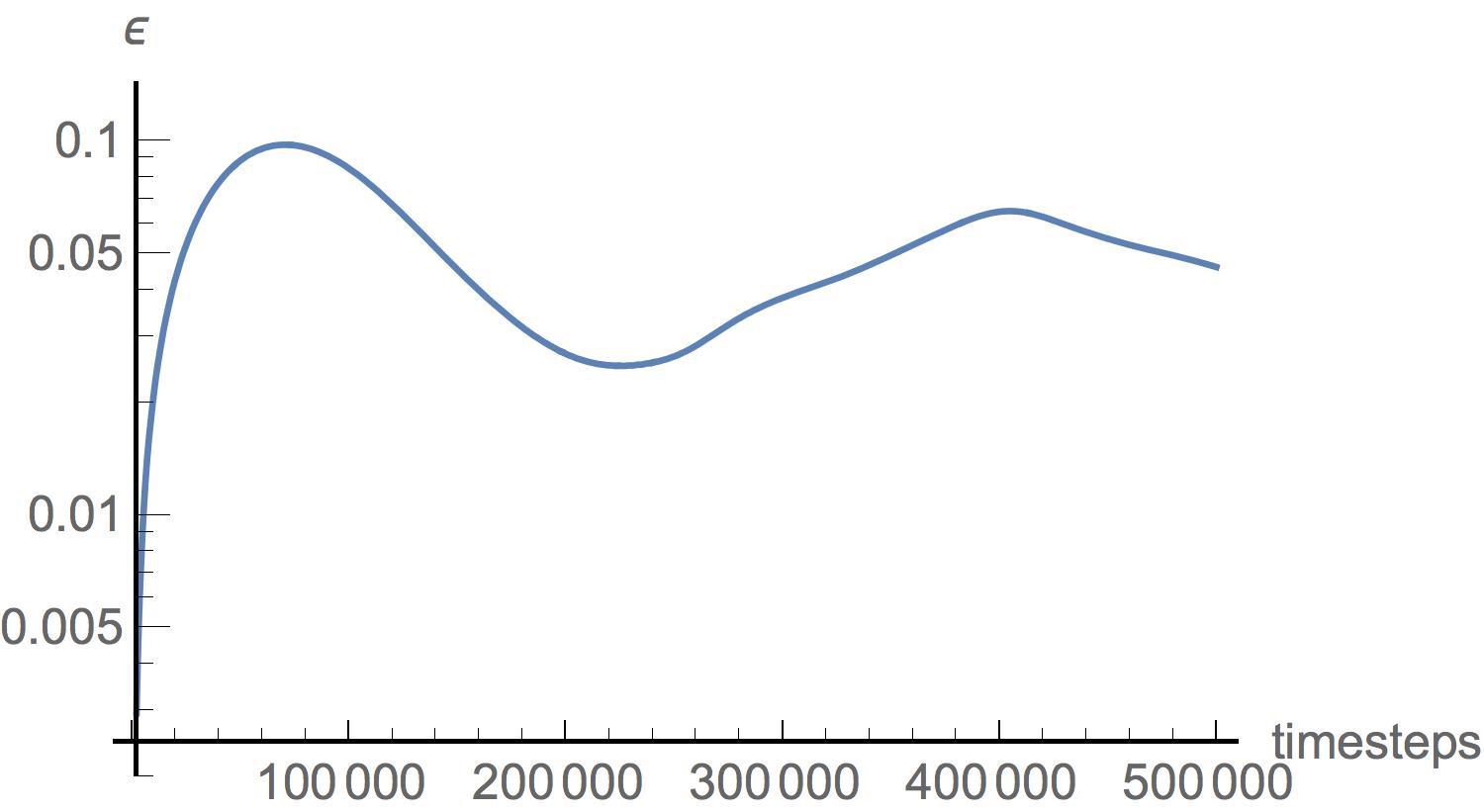}
	\caption[.]{\label{DissPlot}Plot of energy dissipation over time for the Biskamp-Welter profile.}
	\end{figure}

	We now consider the model of Biskamp and Welter \cite{Biskamp1} for decaying 2D MHD turbulence, using their initial profiles
	\begin{eqnarray}
	\vec u(x,y,t=0) = U_0 \left[ \sin(y+0.5) \hat x  - \sin(x+1.4) \hat y \right] \\
	\vec B(x,y,t=0) = B_0 \left[ \sin(y+4.1) \hat x  - 2 \sin(2x+2.3) \hat y \right]
	\end{eqnarray}
	(These are a generalization of the canonical Orszag-Tang vortex).  A snapshot of the current lines are shown in Fig. \ref{BiskampJ} and compared with those from the Biskamp-Welter simulation.
	In Fig. \ref{GammaPlot} we plot the corresponding 2D entropy parameter $\gamma^{*}(x,y)$ at this time snapshot.  The lattice points at which  $\gamma^{*}(x,y) \neq 2$ correspond to points where there are effects of in our partial entropic LB-MHD algorithm. The energy dissipation rate for this Biskamp-Welter case is shown in fig. \ref{DissPlot}. This can be compared with figure 20 in \cite{Biskamp1}.
	
	\section{Conclusion}\label{Sec:Conclusion}
	
	We have extended the Karlin  \cite{entropic1,entropic2} entropic Navier-Stokes algorithm to LB-MHD and tested the ensuing model on 3 different problems:  velocity shear flows exhibiting Kelvin-Helmholtz and/or tearing instability, a generalized Orszag-Tang vortex and magnetized jet instability.  We considered the D2Q9 model for both the particle and vector magnetic field distributions.  The partial entropy algorithm is applied only to the particle distributions while in using a vector distribution for the magnetic field one must allow for magnetic field reversals.  Hence we do not have a fully entropic LB-MHD model.  The algorithm clearly extends immediately to 3D, but because of the much greater computational costs we have restricted our simulations to 2D for we can still capture turbulence effects of the generation of small scale motions since in 2D MHD energy cascades to small scales.  We have found  good agreement with the CFD simulations of Chen \textit{et. al.} and Biskamp and Welter.  The partial entropic algorithm permits much larger ranges of velocity and magnetic field amplitudes than could be found in standard LB-MHD algorithms.  This greater numerical stability is achieved at a quite small increase in computational costs since Karlin \textit{et. al.} have determined a simple algebraic approximation to the full entropic parameter.  This approximation is then carried over as an ansatz for our 2D LB-MHD model.   Moreover the extreme parallelization of this partial entropic LB-MHD algorithm is retained since this algebraic entropic parameter $\gamma^*$ is determined purely from local information at each lattice site. The accuracy of the under-resolved Navier-Stokes simulations of B\"{o}sch \textit{et. al.} \cite{Bosch} portend that this new (partial) entropy method could be a possible subgrid model in itself.  In some sense, this is the spirit behind our pushing the magnitude of $U_0$ and $B_0$.  We are not trying to claim rigorous error bounds on various equilibria representations.  This partial entropic LB-MHD algorithm is a subset of MRT models in which there is now a dynamical relaxation rate determined for quasi-stabilization of the fluid flow
by a well-defined procedure as opposed to the standard static MRT relaxation rates.	
	\section{Acknowledgments}
	This work was supported by an NSF and AFOSR grant.  The computations were performed on DoD Supercomputer \emph{Topaz}.
	
	\bibliography{EntropicBib}
	
\end{document}